\documentclass[onecolumn,12pt, draftclsnofoot, margin=1in]{IEEEtran}

\def\BibTeX{{\rm B\kern-.05em{\sc i\kern-.025em b}\kern-.08em
    T\kern-.1667em\lower.7ex\hbox{E}\kern-.125emX}}

 \usepackage{cite}
\usepackage{amsmath,amssymb,amsfonts, amsthm}
\usepackage{mathtools}
\usepackage{graphicx}
\usepackage{textcomp}
\usepackage{xcolor}
\usepackage[utf8]{inputenc} 
\usepackage{hyperref} 
\usepackage{url} 
\usepackage{booktabs} 
\usepackage{amsfonts} 
\usepackage{nicefrac}
\usepackage{microtype} 
\usepackage{algorithm}
\usepackage[noend]{algpseudocode}
\usepackage{bm, upgreek}
\usepackage{soul}
\usepackage{caption}
\usepackage{wasysym}
\allowdisplaybreaks
\definecolor{red1}{rgb}{0.4,0,0}

\hypersetup{
colorlinks,
 citecolor=blue,
  linkcolor=red1,
  urlcolor=blue
}

\DeclareMathOperator*{\argmax}{arg\,max}

\theoremstyle{plain}
\newtheorem{theorem}{Theorem}[section]
\newtheorem{lemma}[theorem]{Lemma}
 \newtheorem{proposition}[theorem]{Proposition}
\newtheorem{corollary}[theorem]{Corollary}
\newtheorem{definition}[theorem]{Definition}

\newtheorem{assumption}{Assumption} 

\theoremstyle{remark}

\newcommand{\mc}{-\!\!\!\!\circ\!\!\!\!-}
\newcommand{\mathbbm}[1]{\text{\usefont{U}{bbm}{m}{n}#1}}
\newcommand\numberthis{\addtocounter{equation}{1}\tag{\theequation}}
\newcommand{\bR}{\mathbb{R}}
\newcommand{\bE}{\mathbb{E}}
\newcommand{\bZ}{\mathbb{Z}}

\newcommand{\cI}{\mathcal{I}} 
\newcommand{\cO}{\mathcal{O}} 
\newcommand{\FFP}{\mathcal{F}_{1,l}^{(k)}} 

\newcommand{\FSP}{\mathcal{F}_{2,l}^{(k)}}
\newcommand{\IFP}{\mathcal{I}_{1,l}^{(k)}} 

\newcommand{\ISP}{\mathcal{I}_{2,l}^{(k)}}
\newcommand{\FP}{\texttt{P}_1} 
\newcommand{\SP}{\texttt{P}_2} 
 
\newcommand{\iid}{i.i.d.~}
\newcommand{\prob}{ \mathbb{P} }
\newcommand{\cN}{\mathcal{N}}

 \newcommand{\subparagraph}[1]{\textit{#1}} 
\input{tangocolors.sty}

\usepackage{pstricks,pifont}
\usepackage{tikz}
\usepackage{pgfplots,pgfplotstable}
\usepackage{breqn}
\usepackage{amsmath}
\allowdisplaybreaks

\definecolor{red1}{rgb}{0.4,0,0}

\definecolor{red1}{rgb}{0.4,0,0}
\hypersetup{
colorlinks,
 citecolor=blue,
  linkcolor=red1,
  urlcolor=red1
}
\begin{document}
\title{Universal Gaussian Quantization with Side-Information using Polar Lattices}
\author{Shubham~Jha\thanks{Shubham Jha is with the Robert Bosch Center for Cyber-Physical Systems, Indian Institute of Science, Bengaluru 560012, India. Email: \url{shubhamkj@iisc.ac.in}. A preliminary version of this work\cite{Jha20} has appeared in 2020 IEEE Information Theory Workshop (ITW), Italy,  2020. }
}
\maketitle
\date{}
\begin{abstract}
We consider universal quantization with side information
  for Gaussian observations, where the side information
  is a noisy version of the sender's observation
  with  noise variance unknown to the sender.  In this paper, we propose a universally 
  rate optimal and practical quantization scheme for all values of unknown noise variance.
  Our scheme uses
  Polar lattices from prior work, and proceeds based on a structural decomposition  of the underlying auxiliaries so that even when recovery fails
  in a round, the parties agree on a common ``reference point'' that
  is closer than the previous one.  We also present the finite blocklength analysis showing an sub-exponential convergence  for distortion and exponential convergence for rate. The overall complexity of our scheme is $\mathcal{O}(N^2\log^2 N)$ for any target distortion and fixed rate larger than the rate-distortion bound.
\end{abstract}

\section{Introduction}
Distributed quantization with side information at the decoder
is a classic multiterminal information theory problem,
studied first in the seminal work of~\cite{Wyn76}.
We consider  the Gaussian
setting ($cf.$~\cite{Oohama97, stein04, Min15}) where the encoder
and the decoder
observe correlated Gaussian random variables (rvs) $X^N$ and $Y^N$, respectively.
We are interested in the universal version of this problem
where the variance of $X_i$s is known, but the variance
of noise between $X_i$ and $Y_i$ is not known at the encoder. In this setting, we
seek universally rate optimal and practical schemes.

This is indeed a well-studied problem. Perhaps the best
understood variant is where
the encoder observation statistics are unknown
but the channel to the side information is
known; see, for instance,~\cite{Weissman05,Merhav06, Jalali10, Kuzuoka10, Reani11} for results in this setting.
Fewer results are available when the statistics
of channel from $X$ to $Y$ are unknown, which is our setting of interest;
the theory for general sources was studied in~\cite{Watanabe14}
and practical schemes using LDPC codes was considered in~\cite{Dupraz14}.
However, there is still no provably universally rate-optimal, practical code
for this problem. 

Drawing on results on interactive schemes for universal Slepian-Wolf problem
from~\cite{Tyagi18, Banerjee18,Banerjee19}, we propose a practical universal
Wyner-Ziv code between the encoder
and the decoder. Unlike the preliminary version \cite{Jha20} of this work, only encoder is allowed to send messages in the current setting. Our scheme uses Polar lattices from~\cite{Liu19, liu21polartit}, where they were
used for the Gaussian Wyner-Ziv problem with known channel statistics. 
In our scheme,
the encoder communicates its messages in multiple rounds.
In each round, encoder 
considers a new guess from the set of noise variances
and use a code designed for that guess.  We assume that the
set of noise variances used by encoder is predefined and shared upfront with the decoder. That way,
 the decoder exactly knows the round to form the desired estimate. The total message bits communicated to the decoder till this round constitute the overall rate used by our scheme.

As a solution, it is well-known that the encoder needs to sample from an appropriate auxiliary rv $X^\prime$ that forms the following Markov chain structure: $X^\prime \mc X \mc Y.$ However, the joint distribution of $(X^\prime, X, Y)$ depends on the channel statistics between $(X,Y)$-pair, unavailable to the encoder.
To that end, our scheme uses more than one auxiliary and exploit a long resulting Markov chain structure.
By using a structural observation for the underlying auxiliary rvs,
our scheme ensures that even when the guess fails, the encoder and the decoder
agree on a closer ``reference point'' which can be subtracted from both $X^N$
and $Y^N$.  {One of the key property of Polar codes used for the Wyner-Ziv problem (known channel statistics) in  \cite{liu21polartit} is that the polarization operation preserves  degradedness \cite{eren16}.  Further, this property has been instrumental in designing a rateless coding scheme \cite{Li16} based on Polar codes.
 Our scheme too utilizes this property in order to ensure the optimal rate-distortion tradeoff in every round.}
Our presentation below focuses on describing the scheme and presenting the underlying theoretical guarantees that lead to it, which are technical.  Sometimes, we use $\FP$ and $\SP$ to represent {\em encoder} and {\em decoder}, respectively.

\textit{Notation.} Random variables and vectors are denoted in capital letters \textit{without} and \textit{with} bold fonts, respectively. Their realizations are expressed as their small letter counterparts. ${X\sim P}$
implies rv $X$ is distributed as $P$. 
$\mathtt{var}(X)$ denotes the variance of rv $X$, and $\mathtt{cov(\textbf{X})}$ denotes the  covariance of a joint rv $\mathbf{X}$.  $\mathtt{d_{TV}}(P, Q)$ and $\mathtt{D_{KL}}(P\mid\mid Q)$ denote the {\em total variational distance} and {\em Kullback-Liebler (KL)} divergence between two distributions $P$ and $Q$, respectively.

$\mathbf{x}$
is the shorthand for the vector $(x_1, ..., x_n)$, and ${\mathbf{x}(i:j)}$ for
subvector $(x_{i},x_{i+1},..., x_{j})$. $\mathbf{x}_{i:j}$ denotes
the sequence of vectors $\mathbf{x}_{i}, \mathbf{x}_{i+1}, \dots
\mathbf{x}_{j}$. $\mathbf{M}^T$ denotes the transpose of a matrix
$\mathbf{M}$. $\bR$ and $\bZ$ denote the sets of real numbers and integers,
respectively.  $\cO$ represents the standard ``Big O" notation, and we write $f(x)=\cO(g(x))$ as $x\to \infty$ if $\limsup_{x\to \infty} \frac{|f(x)|}{g(x)}<\infty.$

\section{Preliminaries}
\paragraph{The discrete Gaussian distribution}
An $n$-dimensional lattice $\Lambda$ is given by the set  
$\Lambda=\{\lambda=\mathbf{u}\cdot \mathbf{C}: \mathbf{u}\in \bZ^n\},$ where
$\mathbf{C}=[\mathbf{c}_1, \mathbf{c}_2, \dots,
  \mathbf{c}_n]^T$ is a full rank $(n\times n)$-generator matrix.
 {For a vector $\bm{x}\in \bR^n$,  we define the nearest-neighbor quantizer $Q$ associated with $\Lambda$ as $Q_{\Lambda}(\bm{x}):= \text{argmin}_{\lambda\in \Lambda}\|\lambda-\bm{x}\| $
where ties are resolved arbitrarily.  We also define the modulo lattice operation for $\bm{x}$ as $\bm{x}\mod \Lambda := \bm{x} - Q_{\Lambda}(\bm{x}).$}
The joint probability density function (pdf) of an $n$-dimensional Gaussian random vector $\underline{\mathbf{X}}$, with mean  $\bm{\mu}\in \bR^n$ and variance $\sigma^2$  for each independent coordinate, is given by
\begin{align}
  f_{\sigma,\bm{\mu}}(\underline{\mathbf{x}})= \frac{1}{(2\pi\sigma^2)^{n/2}}\exp \left(\frac{\|\underline{\mathbf{x}}-\bm{\mu}\|^2}{2\sigma^2}\right),  \  \underline{\mathbf{x}}\in \bR^n.
\nonumber
\end{align}
Given a lattice $\Lambda$, the {\em discrete Gaussian distribution} over $\Lambda
\subset \bR^n$ centred at $\bm{\mu}$ is defined as
\begin{align}
\mathcal{D}_{\Lambda,\sigma,\bm{\mu}}(\lambda)\triangleq\frac{f_{\sigma,\bm{\mu}}(\lambda)}{\sum \limits_{\lambda'\in \Lambda} f_{\sigma,\bm{\mu}}(\lambda')}, \  \lambda\in \Lambda. \label{dGauss}
\end{align}
Namely, it is a probability mass function (pmf) over points
  of $\Lambda$ with mass of point $\lambda\in \Lambda$ proportional to the Gaussian
  density $f_{\sigma, \bm{\mu}}(\lambda)$ at that point.
Also, define the flatness factor $\varepsilon_\Lambda(\sigma)$~\cite{Cong14} \begin{align*}
\varepsilon_\Lambda(\sigma):= \max \limits_{\mathbf{x}\in \mathcal{R}(\Lambda)}|V(\Lambda)\sum \limits_{\lambda'\in \Lambda} f_{\sigma,\lambda'}(\mathbf{x})-1|, 
\end{align*}
where $\mathcal{R}(\Lambda)$ denotes the fundamental region of $\Lambda$ and $V(\Lambda)$ denotes its volume.
Note that $\varepsilon_\Lambda(\sigma)$ is a decreasing function of $\sigma$, and the normalizing factor in \eqref{dGauss} is bounded as 
\begin{align}\label{e:norm}
\frac{1-\varepsilon_\Lambda(\sigma)}{V(\Lambda)} \leq \sum \limits_{\lambda'\in \Lambda} f_{\sigma,\bm{\mu}}(\lambda')\leq \frac{1+\varepsilon_\Lambda(\sigma)}{V(\Lambda)}.
\end{align}
The following result brings in the importance of flatness factor and shows that the distance between the output distributions for an additive Gaussian channel, when the input is Gaussian and discrete lattice Gaussian, can be controlled using the associated flatness factor of the lattice. In other words,  lattices with small flatness factors can very well approximate the output distribution for any discrete lattice Gaussian input.
  \begin{lemma}[{}\cite{Regev09,Cong14}] \label{l:tvd}
  Consider multivariate rvs $\mathbf{X}, \mathbf{Y},\widetilde{\mathbf{X}},\widetilde{\mathbf{Y}}$ such that
  \begin{align*}
\mathbf{Y}=\mathbf{X}+\mathbf{Z} \ 
\hbox{and }
  \widetilde{\mathbf{Y}}=\widetilde{\mathbf{X}}+\mathbf{Z}, 
  \end{align*}
 where $\mathbf{X}\sim \cN(\bm{0},\sigma_x^2\mathbf{I}_n),  \mathbf{Z}\sim \cN(0,\sigma^2\mathbf{I}_n)$ and $\widetilde{\mathbf{X}}\sim \mathcal{D}_{\Lambda,\sigma_x,\bm{0}}.$ Then, the total variational distance between distributions $P_{\mathbf{y}}$ and $P_{\widetilde{\mathbf{y}}}$ is bounded as
 \begin{align*}
 \mathtt{d}_{\mathtt{TV}} (P_{\mathbf{y}}, P_{\mathbf{\widetilde{y}}})\leq 2\varepsilon,
 \end{align*}
 where $\varepsilon=\varepsilon_{\Lambda}(\frac{\sigma_x\sigma}{\sqrt{\sigma_x^2+\sigma^2}})$ denotes the flatness factor of noise variance scaled by Minimum Mean Square Error (MMSE) coefficient $\sigma_x/\sqrt{\sigma_x^2+\sigma^2}$.
  \end{lemma}

\paragraph{Polar codes}
It will be convenient to recall a general definition of Bhattacharyya
parameter for our discussion on Polar codes.
\begin{definition}
  For a channel $P_{Y|X}$ with a binary input $X$ and (possibly continuous)
output $Y$, the Bhattacharyya parameter $Z(X\mid Y)$ is given by
\[
  Z(X\mid Y)=2\int \limits_y P_Y(y) \sqrt{P_{X|Y}(0|y) P_{X|Y}(1|y)}~dy.
  \]
\end{definition}
The following proposition relates the parameter $Z(X\mid Y)$ with the conditional entropy $H(X\mid Y)$ of the rv $X$ given the rv $Y$.
\begin{proposition}[{\cite[Proposition 2]{Arikan09}}]
For rvs $X$ and $Y$ with $X\in \{0,1\},$ we have
\begin{align}
Z(X\mid Y)^2\leq H(X\mid Y)\leq Z(X\mid Y).\label{BP_bounds}
\end{align}
\end{proposition}
In a Polar code, the input $X^N$ to $N=2^s$ copies of a binary input channel $P_{Y|X}$
is transformed using the {\em generator matrix}
${\mathbf{G}_N=\mathbf{G}^{\otimes s }},$ 
 where ${\mathbf{G}=\begin{bmatrix}
1 & 0\\
1 & 1
\end{bmatrix}}$ and ${\otimes}$ denotes the Kronecker
 product. The transformed bits $U^N=X^N \mathbf{G}_N^{-1}$
under the binary field $\mathbb{F}_2=\{0,1\}$ operations are treated as new inputs, which we try to decode using
 channels $W_i^{(N)}$ from $U_i$
 to $(Y^N, U^{i-1})$. The seminal result of \cite{Arikan09} states
 that the Bhattacharya parameters of the channels $W_i^{(N)}$ tend to
 $0$ or $1$ as $N$ tends to infinity, and the fraction of indices $i$ for which it tends to 0 is exactly the symmetric-capacity of the channel. That is, the channels are ``polarized''
into perfect and useless channels. The indices of the bits with small (close to $0$) Bhattacharyya
  parameters
  constitute the set of {\em information} bits and those with large (close to
  $1$) ones
  constitute the set of {\em frozen} bits.
  The bits indexed in the information set can be determined almost error-free,
  provided that all the bits $U_i$s indexed in the frozen set are shared in advance. In this paper, we will be using Polar codes for {\em degraded channels}, which we define next.
\begin{definition} 
  Consider two channels $W_1 : \mathcal{X} \rightarrow \mathcal{Y}_1$ and $W_2 : \mathcal{X} \rightarrow \mathcal{Y}_2$. The channel $W_1$ is (stochastically) degraded with respect to $W_2$, denoted $W_2 \succeq W_1$, if there exists a channel $V : \mathcal{Y}_2 \rightarrow \mathcal{Y}_1$ such that
 $
W_1(y_1|x)=\int \limits_{y_2\in \mathcal{Y}_2}  W_2(y_2|x)V(y_1|y_2) dy_2. $
\end{definition}
We use the fact that the information set for $W_2$ such that
  $W_2\succeq W_1$ contains the information set for $W_1$ ($cf.$\cite[Lemma 1.8]{Korada09}).
  
{
\paragraph{Polar lattices}
For a pair of lattices $(\Lambda, \Lambda^\prime)$ satisfying $\Lambda^\prime\subset \Lambda, $ $\Lambda^\prime$ is said to be {\em nested} within $\Lambda.$
$\Lambda/ \Lambda^\prime$ denotes the partition of $\Lambda$ into $m:=\frac{V(\Lambda^\prime)}{V(\Lambda)}$ cosets of $\Lambda^\prime$ in $\Lambda$.  We call this as ``binary partition" if $m=2$. Consider a binary partition chain $\Lambda_0/ \Lambda_1/\dots/ \Lambda_\ell$.  For each partition $\Lambda_{i}/\Lambda_{i+1}$, a code $\mathcal{C}_i$ selects a sequence of representatives $a_i\in \{0,1\}$ for the cosets of $\Lambda_{i+1}$. Construction D \cite{Forney2006} requires a set of linear binary codes $\mathcal{C}_1\subseteq \mathcal{C}_2 \cdots \subseteq \mathcal{C}_\ell$. }

{For our problem, we use the Polar lattices \cite{Liu19} which construct capacity achieving Polar codes on each level (based on Construction D) and are known to exhibit a natural nested structure across levels.
It has been shown in \cite{liu21polartit} that Polar lattices have the potential for Gaussian Wyner-Ziv problem, where the solution consists of two nested Polar lattices -- one is AWGN-good and the other is Gaussian rate-distortion bound achieving. This is in accordance with results in \cite{zamir02}, where authors have shown that the Wyner-Ziv problem can be solved by nested quantization-good and AWGN-good lattices. We refer the interested readers to \cite{zamir02, liu21polartit} for more details on the goodness properties of such lattices.
}

\section{Problem Formulation}
We consider the Gaussian rate-distortion problem
where the observations are independent copies
of jointly Gaussian rvs $(X,Y)$ given by 
\begin{align}
X=Y+Z,  \label{modeleq}
\end{align}
where $Y$ and $Z$ are independent
Gaussian rvs with zero means
and variances $\sigma_{y}^2$ and $\sigma_{z}^2$,
respectively.  
For our setting,  it is more convenient to fix the
variance $\sigma_x^2$ of $X$ and express $\sigma_{y}^2$
as $\sigma_x^2-\sigma_{z}^2$.

Specifically,  let $\{(X_i, Y_i)\}_{i=1}^N$ be \textit{N}
\emph{independent and identically distributed} (i.i.d.) copies of Gaussian rvs with joint pdf
$P_{XY}=\cN\left([
0~0]^\top,  \mathbf{K}\right),
$
where for $\sigma_x^2> \sigma_{z}^2, \mathbf{K}$ is the covariance matrix given by $\left[\begin{matrix}
\sigma_x^2 & \sigma_x^2-\sigma_{z}^2 \\
\sigma_x^2-\sigma_{z}^2 & \sigma_x^2-\sigma_{z}^2
\end{matrix}\right].$
For brevity, we use the abbreviation $(\mathbf{X}, \mathbf{Y}):=
\{(X_i, Y_i)\}_{i=1}^N$.  Throughout this paper, for an estimate $\hat{\mathbf{x}}$ of any vector $\mathbf{x},$ we fix the {\em distortion measure} $D$ to be the squared Euclidean distance  given by 
$D(\mathbf{x}, \mathbf{\hat{x}}):=\|\mathbf{x}-\hat{\mathbf{x}}\|^2$.  

While there can be various possible applications,  our formulation is guided by the following application.  Suppose parties $\FP$ and $\SP$ have access to two correlated files $F_1$ and $F_2$, respectively,  such that the ``amount'' of correlation between files is known only to $\SP$.
$\FP$ observes $F_1$ and prepares a compressed version comprising multiple small fragments.
{On the other hand, $\SP$ observes $F_2$ and uses its knowledge of correlation between $F_1$ and $F_2$
to download as few number of compressed segments as needed to recover $F_1$ to a prescribed distortion using these compressed fragments.}
The challenge here is that $\FP$ is not aware of the correlation between files, which makes it difficult to compress $F_1$ appropriately.  When $\FP$ knows the correlation,  it can compress $F_1$ using standard Wyner-Ziv codes. However, in the absence of this knowledge, we need a universal
coding scheme.We capture the requirement above formally as follows: 

Parties $\FP$ and $\SP$ observe the sequences $\mathbf{X}$ and $\mathbf{Y}$, respectively,  generated according to $P_{XY}$, and the goal for $\SP$ is to estimate $\FP$'s observation $\mathbf{X}$ within a fixed target distortion $\Delta$. We assume that number of samples observed is large enough such that parties can infer the marginal moments upto an acceptable accuracy. Further, to model the nescience of correlation and $\mathbf{Y}$'s uncertainty at $\FP$, and thereby,  to capture the universal behaviour, we make the following assumption.
\begin{assumption}\label{a:1}
The variance $\sigma_x^2$ is known to both $\FP$ and $\SP$, but $\sigma_{z}^2$ is known only to $\SP$. Further, $\sigma_z^2$ lies in a closed positive interval $\cI\subseteq \bR_{+}.$
\end{assumption} 
We consider schemes where $\FP$ encodes $\mathbf{X}$ using a finite sequence of $r$ increasing rates $R_1, \dots, R_r$ representing $r$ different fragments of encoded data. $\SP$ downloads the first $k$ segments of total rate $R_1 + ... + R_k$, where $k$ is decided by $\SP$ using its knowledge of
the correlation $\sigma_z^2$. 

More formally, we consider $r$-round Wyner-Ziv (WZ) codes
consisting of encoders and decoders $(e_i, d_i), i\in\{1,\ldots,r\}$. Each $e_i$ is an encoder of rate $R_i$ whose output $C_i$, given by
$C_i=e_{i}(\mathbf{X}),$ is an $NR_i$ length bit-string
and each $d_i$ is a decoder that uses $C_1,\dots, C_i$ along with the side-information $\mathbf{Y}$ to form an estimate of $\mathbf{X}$. $\SP$ forms the estimate $\hat{\mathbf{X}}$ in any round $k$ by applying the decoder $d_k$ given by $\hat{\mathbf{X}}=d_k(C_{1}, ..., C_{k}, \mathbf{Y})$. 
Note that $d_k$ may use previous decoder outputs till round $k-1$.


 
Recall that when $\sigma_z^2$ is known at both encoder and decoder, for $N$ sufficiently large,
the minimum rate $R^\ast$ required to attain distortion 
$\Delta$ is roughly $\frac{1}{2} \log \frac{\sigma_{z}^2}{\Delta}$ \cite{liu21polartit,Wyn76}.
Our goal is to design codes attaining this rate {\em universally} (that is, even when $\FP$ doesn't know $\sigma_z^2$) for all values $\sigma_{z}^2$, which are known to be {lying in a fixed interval $\cI$}.
Specifically, the universal $r$-round WZ code definition below requires that for all (unknown) noise variances $\sigma_z^2$, there must exists a integer $k\leq r$ such that
$d_k$ can recover an estimate of $\mathbf{X}$
from $(C_1, ..., C_k, \mathbf{Y})$ and the total rate used $R_1 +...+R_k$
is roughly $\frac{1}{2} \log \frac{\sigma_{z}^2}{\Delta}$ (the optimal rate for known $\sigma_z^2$).

\begin{definition}[Universal WZ codes]\label{d:univ}
  For $\delta,\epsilon >0$, a fixed
  $\Delta>0$ and a closed interval $\cI\subset \bR_+$, 
  an $r$-round WZ code is {\em $(\epsilon, \delta)$-universal} at
    distortion level $\Delta$ for $\cI$ if for every 
     $\sigma_{z}^2\in \cI$, there exists a $k\leq r$ such that
     \[\sum_{j=1}^{k} R_{j} \leq \frac 12 \log \frac  {\sigma_{z}^2}{\Delta}+\epsilon, \ \  \bE \|\mathbf{X}- \mathbf{\hat{X}}\|^2 \leq N\Delta+\delta.\]
\end{definition}
We emphasize that we don't consider the related problem of identifying the appropriate $k$ using $\mathbf{X}$ and $\mathbf{Y}.$ A particular method for this,which requires $\SP$ to form an estimate $\hat{\mathbf{X}}$ and compare it with $\mathbf{X}$ using its hash, was considered in an earlier version of this paper \cite{Jha20}. In the current setting, $\SP$ decides the right $k$ in the beginning itself based on the knowledge  $\sigma_z^2.$ We describe this later in next section.

It is important to note that for $\sigma_{z}^2 \leq \Delta$, the estimate $\hat{\mathbf{X}}=\mathbf{Y}$ constitutes an acceptable estimate, and therefore, we are interested in the case when $\sigma_{z}^2 \geq \Delta.$ {Accordingly,  we assume that $\omega\geq \Delta$,  for all $\omega\in \cI$.}

{\em A remark on terminology:} For consistency with the earlier conference version of the paper, which was addressing a slightly different interactive variant of the problem, we will use the phrase ``$k$-round code'' to represent the code corresponding to encoders $e_1, ..., e_k$ and decoder $d_k$. 


\section{Proposed Scheme for Universal Quantization}
In this section,  we propose the strategies for $\FP$ and $\SP$ achieving the rate-distortion bound universally.  
\subsection{A review of the basic Polar code based scheme}
We will review first the classic scheme from \cite{Wyn76}, which forms the basis of many practical schemes. In that setting,  $\sigma_{z}^2$ is assumed to be known at the encoder $\FP$ too.
The scheme uses an auxiliary rv $X^\prime$,  which minimizes the conditional mutual-information $I(X\wedge X^\prime \mid Y)$ and is independent of $Y$ given
$X$.  For the Gaussian case which is of interest to us,
this auxiliary takes a simple form given by ($cf.$~\cite{Oohama97})
\begin{equation}
X^\prime=X+T, \label{mc1}
\end{equation}
where $T$ is a Gaussian rv with mean zero and variance
$
\bar{\Delta}=\frac{\sigma_{z}^2\Delta}{\sigma_{z}^2-\Delta},
$
  independent of $(X,Y)$.
  {Denote by $P_{X^\prime XY}$ the joint distribution of rvs due to \eqref{modeleq} and \eqref{mc1}.  Given the pair of rvs $(X^\prime, Y)$, one can form the MMSE estimate $\hat{X}_{MMSE}$ of $X$ given by $\hat{X}_{MMSE}:=\bE_{P_{X\mid X^\prime Y}}[X\mid X^\prime,Y],$ for which the MSE is $\Delta.$ Then,} a naive solution for $\FP$ is to generate  $N$ independent samples $X^{\prime N}$ from this auxiliary such that $(X^\prime_{t},X_t, Y_{t})_{t=1}^N$  are i.i.d. and send them
to $\SP$, who in turn, uses $X^{\prime N}$ and $Y^N$ to construct $\hat{X}_{MMSE}$.
However, this will require too much communication.  To alleviate that,  one can use shared randomness to simulate
these samples $X^{\prime N}$ at $\FP$ and send them to $\SP$ using much less communication.
This is an interpretation of the classic Wyner-Ziv scheme; the scheme in~\cite{Liu19}, too, can be interpreted in this manner.

To facilitate the simulation mentioned above, it is more appropriate
to consider an alternative form of the Markov model in~\eqref{mc1}.
Specifically, we consider the following generative model:
\begin{align}
X=\frac{\sigma_x^2}{\sigma_x^2+\bar{\Delta}}X^\prime+ T^\prime \text{ and } \bar{Y}=X+Z^\prime, \label{mc2}
\end{align}
where $T^\prime$ and $Z^\prime$ are independent Gaussians with zero
means and variances $
\frac{\sigma_x^2\bar{\Delta}}{\sigma_x^2+\bar{\Delta}}$ and $\frac{\sigma_x^2\sigma_{z}^2}{\sigma_x^2-\sigma_{z}^2}$, respectively.
{Let $A:=\frac{\sigma_x^2}{\sigma_x^2+\bar{\Delta}}X^\prime$, and denote by $Q_{X^\prime XY}$ the joint distribution of rvs in \eqref{mc2}.}
Then, the above generative model satisfies $P_{X^\prime XY}=Q_{X^\prime XY}$,  and the pair $(A,\bar{Y})$ still allows us to form an estimate of $X$ which is as accurate
as that can be formed using $(X^\prime,Y)$.  {In particular, the corresponding MMSE estimate $\bE_{Q_{X\mid X^\prime Y}}[X\mid X^\prime,Y]=\hat{X}_{MMSE}$ and the MSE value is $\Delta$.} Also, the mutual information $I(A\wedge X \mid \bar{Y})$
equals $\frac 1 2 \log\frac {\sigma_{z}^2} \Delta$, the optimal rate for getting distortion $\Delta$.

However, the following problem still remains. We need to quantize the samples before communicating. Towards that, several quantization methods 
have been proposed using structured codes of which the most recent,  \cite{Liu19}, is using Polar codes from~\cite{Arikan09}. 
  The idea is to use a lattice Gaussian rv from~\eqref{dGauss} instead of a continuous one. In particular,  a discrete Gaussian rv $\widetilde{A}$ is considered  
over a one-dimensional lattice $\Lambda=N^{-1/2}\cdot\bZ$  instead of the Gaussian rv $A$, and \eqref{mc2} is modified as
  \[\widetilde{X}=\widetilde{A}+ T^\prime \hbox{ and }
  \widetilde{Y}=\widetilde{X}+Z^\prime.\] Note that the choice for $\Lambda$ is not arbitrary.  It was shown in \cite{liu21polartit} that for $c=\cO(N^{-1/2})$,  the ``flatness factor'' associated with the lattice $c\cdot \bZ$ is negligible, which further ensures that the induced distribution $P_{\widetilde{X}\widetilde{Y}}$ is close to $P_{XY}$ in total variational distance ($cf.$ Lemma \ref{l:tvd}).  For the sake of completeness, we present this in Proposition \ref{prop3}.  Further,  to ease our presentation,  we take the orderwise constant {in the value of $c$} to be unity.  Note that while rv $\widetilde{A}$ takes values only in the lattice $N^{-1/2}\cdot \bZ$,  rvs $T^\prime$ and $Z^\prime$ take values in $\bR$. 
    
    Further, due to this closeness in joint pdfs,  the samples $X^N$ to be quantized can be approximated as $N$ independent copies
$\widetilde{X}^N$ of $\widetilde{X}$. This is tantamount to viewing
$X^N$ as being generated by first generating $\widetilde{A}^N$ and then adding Gaussian noise to it.  From here on, we will simply view our observations as coming from this new modified
  distribution.

The rest of the scheme proceeds as before, and the parties
use structured codes to simulate
$\widetilde{A}^N$. However, this new auxiliary is still an infinite-precision
number. The last component of lattice construction in~\cite{Liu19} is 
the observation that we need not recover $\widetilde{A}_i$s completely, and
it suffices to agree on the $\ell$ least significant bits with $\ell=\mathcal{O}(\log N)$.  It is useful to note that this choice of $\ell$ together with the lattice $N^{-1/2}\cdot\bZ$ has been crucial in establishing a sub-exponential convergence to the optimal rate-distortion bound $R^\ast (\Delta)$.   

Finally, the scheme uses Polar codes to
simulate and share 
the $\ell$ least significant bits of each coordinate of $\widetilde{A}^N$
at $\SP$.  Specifically, $\FP$ uses Polar codes as a covering code
to recover the information bits of $\widetilde{A}^N$ at each level $1\leq j \leq \ell$,  and $\SP$ uses it as a packing code for the channel from $\widetilde{A}$ to $\widetilde{Y}$;
the common frozen bits are sampled from shared randomness. 
\subsection{The universal scheme}
Coming to our universal case, since $\sigma_{z}^2$ is not known to $\FP$, it cannot
fix the distribution of $\widetilde{A}$ upfront. Instead, we consider $r$ distinct auxiliaries
for our scheme motivated by the infinite divisibility property of Gaussians.  Each of these auxiliaries corresponds to a different possible
value of the unknown $\sigma_{z}^2$.  Without loss of generality, let $\cI$ ($c.f.$ Definition \ref{d:univ}) takes the form $\cI:=[\sigma_0^2, \sigma_r^2]$ for some $\sigma_0, \sigma_r > 0.$
Further,  consider a finite grid of points to cover the entire continuum $\cI$. In particular,  let $\sigma_0^2\leq \dots \leq \sigma_{r}^2$ be an increasing $r$-tuple partitioning $\cI$ into $r$ sub-intervals.  We assume that the tuple $(\sigma_0^2, \dots, \sigma_{r}^2)$ is known to both $\FP$ and $\SP$.  For $1\leq k\leq r, $ the value $\sigma_k^2$ corresponds to the possibility that
$\sigma_{z}\in[\sigma_{k-1}, \sigma_{k})$.

\begin{figure}[t]
\centering \includegraphics[scale=0.4]{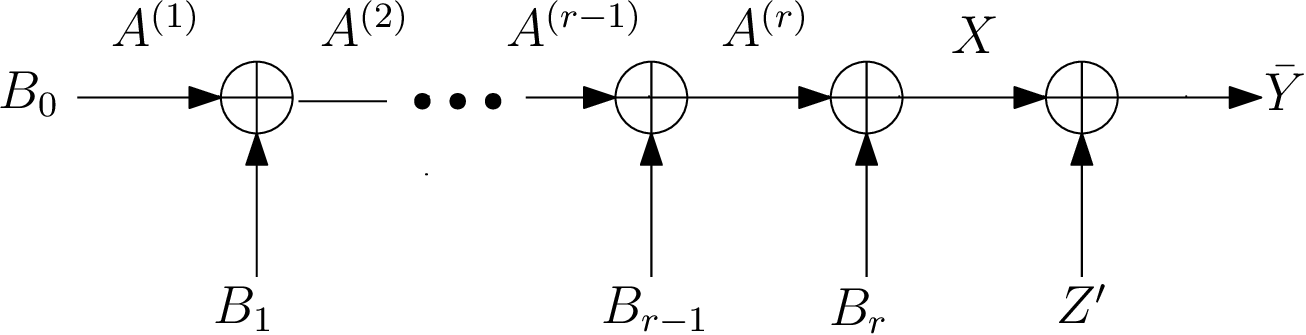}
\caption{General Markov-chain for $r$ round WZ}
\label{generalmc}
\end{figure}

Denote by  $A^{(1)}, ..., A^{(r)}$ the optimal auxiliaries corresponding to noise variances $\sigma_0^2,\dots, \sigma_{r}^2,$ respectively. 
 These auxiliaries can be viewed as forming a Markov chain
  depicted in Fig. \ref{generalmc}.
This Markov chain which couples all these auxiliaries is instrumental in the design of our scheme. 
Specifically, we observe that the
  auxiliary $A^{(k)}$ used in round $k$ can be decomposed
\begin{align}\label{d:aux}
A^{(k)}=\sum\limits_{j=0}^{k-1}B_j,
\end{align}
 where rvs
$B_j$ are specified by following sequence of distributions: 
\begin{align*}
B_0&\sim \cN(0, \alpha_1\sigma_x^2),\\
B_{i}&\sim\mathcal{N}\left(0, \alpha_{i}\Delta_i-\alpha_{i+1}\Delta_{i+1}\right),  \ 1\leq i\leq r-1, \hbox{ and }\\
B_{r} &\sim\mathcal{N}\left(0, \alpha_{r}\Delta_r\right), 
\end{align*}
with
$\Delta_i=\frac{\sigma_i^2\Delta}{\sigma_i^2-\Delta} \hbox{ and } \alpha_{i}=\frac{\sigma_x^2}{\sigma_x^2+\Delta_i}, ~\forall i{\in} [r]
.$
That implies, for $1\leq k\leq r$,  $A^{(k)}$ is a Gaussian rv with mean zero and variance $\alpha_{k}\sigma_x^2.$  
This decomposition is the key step in designing our rate-optimal  strategy.

 In round $k$, we subtract
  the previously recovered ``parts'' $B_0, ..., B_{k-2}$ and treat
  the residue as the new observation, i.e.,  the pair $(X, \bar{Y})$ is replaced by $(X-A^{(k-1)}, \bar{Y}-A^{(k-1)})$. 
  The main idea driving our scheme
  is that even when
$\sigma_{z}^2\geq \sigma_{k}^2$,
  both $\FP$ and $\SP$ will end-up recovering $B_{k-1}$\footnote{Since $\mathtt{P}_2$ has the knowledge of both the noise variances $\sigma_k^2$, used by $\mathtt{P}_1$ to construct $B_{k-1}$, and $\sigma_z^2$, it perfectly knows the channel variance $\mathtt{var}(T_k+Z^\prime)$ required to form the decoder for $B_{k-1}$.}, and thereby $A^{(k)}$, which is an optimal auxiliary for the noise variance $\sigma_k^2.$
In round $k+1$, this can be subtracted from both $X$ and $\bar{Y}$ by $\FP$ and $\SP$, respectively.

\begin{figure}
 \centering  \includegraphics[scale=0.4]{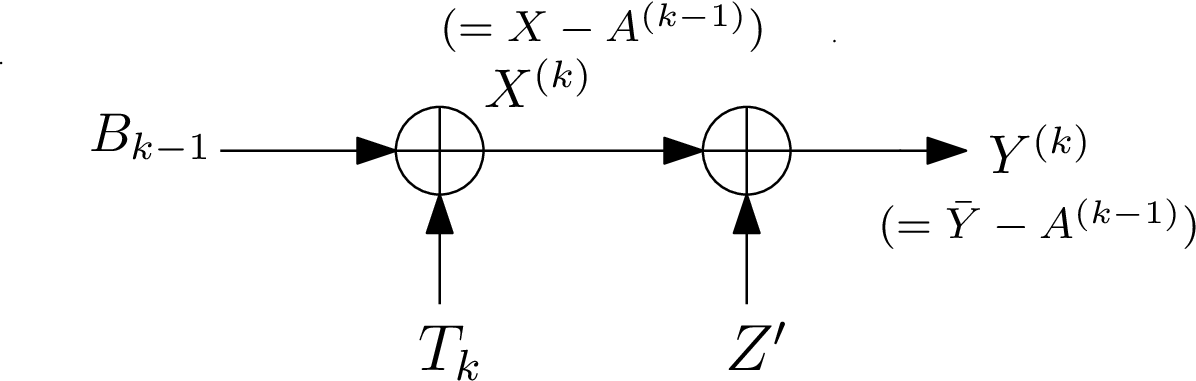} 
\caption{Markov chain in round $k$ }
\label{k-rnd-ideal}
\end{figure}

   Heuristically, when the parties begin, they only agree on the origin $0$
  as the ``reference point''. But in each round they agree on a (on average)
  closer
  reference point, which they subtract from both their observations.
Since,  $\SP$ knows the value of $\sigma_z^2$, it exactly knows {the number of rounds $k$ needed for its estimation task. } 		
The new observation pairs for round $k,$ denoted as $(X^{(k)}, Y^{(k)}),$ is obtained by subtracting $A^{(k-1)}$ from the pair $(X,\bar{Y}).$ 
We illustrate this distribution in Fig.~\ref{k-rnd-ideal}.
Note that the Markov chain for round $k$:
 \begin{equation}
X^{(k)}=B_{k-1}+T_k \text{ and } Y^{(k)}=X^{(k)}+Z', \label{mc3}
 \end{equation}
 where ${T_{k}}=\sum\limits_{i=k}^rB_i$ has distribution
 $\mathcal{N}(0,\alpha_{k}\Delta_{k})$.
For consistency, we take $(X^{(1)},Y^{(1)})=(X,\bar{Y})$.
Further, as described earlier, we simulate samples from the lattice Gaussian distribution 
  $\widetilde{B}_{k-1}\sim\mathcal{D}_{N^{-1/2}\bZ, \mathtt{var}(B_{k-1})},\bm{0}$ instead of
  $B_{k-1}$ and modify \eqref{mc3} as 
\begin{equation}
\widetilde{X}^{(k)}=\widetilde{B}_{k-1}+T_k\text{ and }\widetilde{Y}^{(k)}=\widetilde{X}_k+Z'. \label{mc4}
\end{equation}
Henceforth, we consider $N$ \iid samples from the distribution in
 \eqref{mc4}, $i.e.$,
\begin{equation}
\widetilde{\mathbf{X}}^{(k)}=\widetilde{\mathbf{B}}_{k-1}+\mathbf{T}_k,\text{ and } \widetilde{\mathbf{Y}}^{(k)}=\widetilde{\mathbf{X}}^{(k)}+\mathbf{Z}'. \nonumber
\end{equation}  
For $1\leq l \leq \ell$, let
${\mathbf{U}_l^{(k)}\triangleq\mathbf{H}_l\textbf{G}_N^{-1}}$ where
each coordinate $\mathbf{H}_l(j)$ of $\mathbf{H}_l$ corresponds to the $l$-th least significant bit in the binary representation of scaled $j$-th coordinate
$N^{1/2}\widetilde{\mathbf{B}}_{k-1}(j)\in \bZ.$
Recall that $\mathbf{G}_N$ is the $N\times N$ generator matrix for Polar codes. It has been noted in \cite{Liu19, liu21polartit} that $\forall l\in \{1,\dots,\ell\}$, the channel between $\mathbf{H}_l$ and $\widetilde{\mathbf{X}}^{(k)}$ conditioned on the event $\{\mathbf{H}_{1:l-1}=\mathbf{h}_{1:l-1}\}$ 
may not be \emph{symmetric} in general.
For 
$\FP$, the index set $[N]$ is partitioned into the
information set
$\IFP$ and the frozen set $\FFP$ 
defined as follows:  
$\FFP$ is the set of indices $j\in [N]$ satisfying for any $\beta\in (0,1/2)$,
\begin{align*}
Z\left(\mathbf{U}_l^{(k)}(j)\mid
  \mathbf{U}_l^{(k)}(1:j-1),\mathbf{U}_{1:l-1}^{(k)},
  \widetilde{\mathbf{X}}^{(k)} \right) &\geq 1-2^{-N^\beta}, \text{or } \nonumber \\
  Z\left(\mathbf{U}_l^{(k)}(j)\mid
  \mathbf{U}_l^{(k)}(1:j-1), \mathbf{U}_{1:l-1}^{(k)}\right) &\leq
  2^{-N^\beta}.\numberthis \label{e:bhat_cc2}
\end{align*}
Similarly,  the index set $[N]$ is partitioned into
$\ISP$ and $\FSP$.
$\ISP$ is the set of indices $j\in [N]$ satisfying, for any $\beta\in (0,1/2)$,
\begin{align}
Z\left(\mathbf{U}_l^{(k)}(j)\mid
  \mathbf{U}_l^{(k)}{(1:j-1)},\mathbf{U}_{1:l-1}^{(k)},
  \widetilde{\mathbf{Y}}^{(k)} \right)&\leq 2^{-N^\beta}, \text{ and }
\nonumber
  \\ 
Z\left(\mathbf{U}_l^{(k)}(j)\mid
  \mathbf{U}_l^{(k)}(1:j-1), \mathbf{U}_{1:l-1}^{(k)}\right)&\geq
  1-2^{-N^\beta}.
  \nonumber
\end{align}
These definitions of information and frozen sets are from~\cite{yamamoto13} where Polar codes for asymmetric channels
    were analysed. It has a slightly different form in comparison to the original
  definition in~\cite{Arikan09}.
Note that we have defined the frozen set for $\FP$ and the information set for $\SP$. The reason for this distinction is that we use Polar codes to construct a covering (source) code for $\FP$ and a packing (channel) code for $\SP$; see~\cite{yamamoto13}.  Since the channel between  each $\mathbf{H}_l, $ $1\leq l\leq \ell$,  and $\mathbf{Y}_{k}$ is perfectly known to the decoder,  it constructs its frozen set $\FSP$ in advance for all the levels and shares them with $\FP$. 
\begin{algorithm}[t]
\caption{$\FP$'s strategy in round $k$}
\textbf{Require:} $\sigma_k^2$,  $\Delta$, $\Lambda=N^{-1/2}\bZ$,  $\ell$,  $\{d\mathcal{F}_{l}^{(k)}\}_{l\in [\ell]}, \mathbf{x}^{(1)}=\mathbf{x}, \mathbf{a}^{(k-1)},$ shared randomness  \eqref{shared_rand}\\ 
\textbf{Initialize:} $\mathbf{x}^{(k)}\leftarrow \mathbf{x}-\mathbf{a}^{(k-1)}, \mathbf{u}_{1:\ell}^{(k)}\leftarrow [\bm{0}]_{N\times\ell}$
\begin{algorithmic}[1]
\For {$l \in [\ell]$}
\vspace{2pt}
\For {$j \in [N]$}
\If {$j \in \IFP$} 
\State Compute $p_{1,l}^{(k)}(j)$ as in~\eqref{post_prob}
\State Set $\mathbf{u}_l^{(k)}(j) \leftarrow
\mathtt{Bernoulli}(1/1+p_{1,l}^{(k)}(j))
$ \label{line5a1}
\Else~Use \eqref{shared_rand} to determine $\mathbf{u}_l^{(k)}(j)$
\EndIf
\EndFor
\State Send $\mathbf{u}_l^{(k)}(d\mathcal{F}_{l}^{(k)})$ to $\SP$ \label{line9a1}
\Comment{ \texttt{$d\mathcal{F}_{i,l}^{(k)}$: set difference of frozen-bit sets}}
\EndFor
\State $\mathbf{b}_{k-1} \leftarrow N^{-1/2}\left((\mathbf{u}_{1}^{(k)}+...+2^{\ell-1}\mathbf{u}_{\ell}^{(k)})\mathbf{G}_N~\text{mod } 2^\ell \bZ\right)$\Comment{\texttt{Lattice point}}
\State $\mathbf{a}^{(k)}\leftarrow \mathbf{a}^{(k-1)}+\mathbf{b}_{k-1}$\Comment{ \texttt{Auxiliary update}}
\end{algorithmic}
\label{algo1}
\end{algorithm}

\begin{algorithm}[t]
\caption{$\SP$'s strategy in round $k$}
\textbf{Require:} $\sigma_k^2$,  $\Delta$, $\Lambda=N^{-1/2}\bZ$,  $\ell$, $\{d\mathcal{F}_{l}^{(k)}\}_{l\in [\ell]}, \mathbf{\bar{y}}=\frac{\sigma_x^2}{\sigma_x^2-\sigma_{k}^2}\cdot \textbf{y},$ $\mathbf{\hat{a}}^{(k-1)}$, shared randomness \eqref{shared_rand}\\ 
\textbf{Communication received:} $\left\{\mathbf{u}_l^{(k)}(d\mathcal{F}_{l}^{(k)})\right\}_{l\in [\ell]}$\\
\textbf{Initialize:} $\mathbf{y}^{(k)}\leftarrow \mathbf{\bar{y}}-\mathbf{\hat{a}}^{(k-1)}, \mathbf{\hat{u}}_{1:\ell}^{(k)}\leftarrow [\bm{0}]_{N\times \ell}$
\begin{algorithmic}[1]
\For {$l \in [\ell]$}
\vspace{2pt}
\For {$j \in [N]$}
\If {$j \in \ISP$}
\State Compute $p_{2,l}^{(k)}(j_u)$ as in \eqref{post_prob2}
\State $\mathbf{\hat{u}}_l^{(k)}(j)\leftarrow \underset{u\in \{0,1\}}\argmax\ p_{2,l}^{(k)}(j_u) $ \Comment{ \texttt{MAP rule}}\label{line5a2}
\ElsIf {$j \in d\mathcal{F}_{l}^{(k)}$} \vspace{2pt}
\State $\mathbf{\hat{u}}_l^{(k)}(j)\leftarrow\mathbf{u}_l^{(k)}(j)$ 
\Comment {\texttt{Received bits}}
\Else{} Use \eqref{shared_rand} to determine $\mathbf{\hat{u}}_l^{(k)}(j)$
\EndIf
\EndFor
\EndFor
\State $\mathbf{\hat{b}}_{k-1}\leftarrow N^{-1/2}\left((\mathbf{\hat{u}}_1^{(k)}+...+2^{\ell-1}\mathbf{\hat{u}}_{\ell}^{(k)})\mathbf{G}_N~\text{mod } 2^\ell \bZ\right)$
\Comment {\texttt{Lattice point}}
\vspace{3pt}
\State $\mathbf{\hat{a}}^{(k)}\leftarrow \mathbf{\hat{a}}^{(k-1)}+\mathbf{\hat{b}}_{k-1}$
\Comment{ \texttt{Auxiliary update}}
\vspace{3pt}
\State $\mathbf{\hat{x}}^{(k)} \leftarrow \mathbf{\hat{a}}^{(k)}+\frac{\sigma_k^2\Delta(\sigma_x^2-\sigma_z^2)}{\sigma_x^2\Delta(\sigma_k^2-\sigma_z^2)+\sigma_z^2\sigma_k^2\sigma_x^2}\left(\mathbf{\bar{y}}- \mathbf{\hat{a}}^{(k)}\right)$ 
\Comment{ \texttt{Reconstruction}}
\end{algorithmic}
\label{algo2}
\end{algorithm}
With that, we specify the overall protocol to be
  used in round $k$ by $\FP$ and $\SP$
  in Algorithms \ref{algo1} and \ref{algo2}, respectively.
$\SP$ uses MAP rule for realizing its information bits with posterior $p_{2,l}^{(k)}(j_u)$ given by 
\begin{align*}
p_{2,l}^{(k)}(j_u)&=P_{\mathbf{\hat{U}}_l^{(k)}(j)|\mathbf{\hat{U}}_l^{(k)}(1:j-1) \mathbf{\hat{U}}_{1:l-1}^{(k)}\mathbf{Y}^{(k)}} (u\mid\mathbf{\hat{u}}_l^{(k)}(1:j-1), \mathbf{\hat{u}}_{1:l-1}^{(k)}, \mathbf{y}^{(k)}),  \numberthis \label{post_prob2} 
\end{align*}
where as 
$\FP$ uses randomized MAP rule with quantity $p_{1,l}^{(k)}(j)$ defined as ratio
\begin{align}
\frac{P_{\mathbf{U}_l^{(k)}(j)|\mathbf{U}_l^{(k)}(1:j-1)\mathbf{U}_{1:l-1}^{(k)}\mathbf{X}^{(k)}}(0\mid\mathbf{u}_l^{(k)}(1:j-1), \mathbf{u}_{1:l-1}^{(k)},\mathbf{x}^{(k)})}{P_{\mathbf{U}_l^{(k)}(j)|\mathbf{U}_l^{(k)}(1:j-1)\mathbf{U}_{1:l-1}^{(k)}\mathbf{X}^{(k)}}(1\mid\mathbf{u}_l^{(k)}(1:j-1), \mathbf{u}_{1:l-1}^{(k)},\mathbf{x}^{(k)})} \label{post_prob}
\end{align}
for the same. The paper \cite{Korada10} discusses several
advantages of using a randomized MAP rule while encoding.

Also, due to 
a (stochastically) degraded channel structure between the channels
from $\mathbf{H}_l$ to $\widetilde{X}_k$ and $\mathbf{H}_l$ to $\widetilde{Y}_{k}$ conditioned on rvs $\mathbf{H}_{1:l-1}$,  we notice that  $\FFP \subseteq \FSP~ \forall l\geq 1$. Therefore, $\FP$ and $\SP$
use randomized mapping to realize the bits in all the levels 
$\{\FFP\}_{l\geq 1}$, while remaining bits in
$\{d\mathcal{F}_{l}^{(k)}:=\FSP \setminus \FFP\}_{l\geq 1}$ are communicated as $C_{k}$.  This
randomized mapping can be shared in advance and 
realized using pseudo-random numbers generated as follows (see \cite{yamamoto13}):  For all $l\geq 1$, $j\in [N]$,
\begin{align}\label{shared_rand}
\mathbf{u}_l^{(k)}(j)&\sim \mathtt{Ber}(P_{\mathbf{U}_l^{(k)}(j)|\mathbf{U}_l^{(k)}(1:j-1)\mathbf{U}_{1:l-1}^{(k)}}(1\mid\mathbf{u}_l^{(k)}(1:j-1), \mathbf{u}_{1:l-1}^{(k)}))
\end{align}
Once $\SP$ decodes all its frozen bits located in $\FSP$ using communicated bits $C_{k}$ and the shared randomness,  the remaining bits  in $\ISP$ are recovered with high probability using MAP rule (See Line \ref{line5a2},  Algorithm \ref{algo2}). This is essentially due to the capacity-achieving property of Polar codes.  The final step in  Algorithm \ref{algo2} is to add all the previously recovered lattice points (auxiliaries) and form the MMSE estimate.
%

\section{Analysis}\label{sec5}
Our analysis can be understood by first considering
  the Gaussian distribution in~\eqref{mc3}, and then moving, in steps,
  to the discrete Gaussian distributions $\{\widetilde{A}^{(i)}: i\in [r]\}$, and finally
  to the distribution simulated using Polar codes, retaining only the $\ell$
  least significant bits.
Following~\cite{liu21polartit},  we show that all these distributions
are close to each other and mean-squared error guarantees for one translates to that for the other.
Indeed, following a result from~\cite{Cong14}, we notice that for the discrete
Gaussian auxiliary $\widetilde{A}^{(i)}$ taking $N$ to be sufficiently large
renders $P_{\widetilde{X}_i\widetilde{Y}_i}$ close to $P_{X_iY_i}$.  Further,
a covering bound for Polar codes for asymmetric channels from~\cite{yamamoto13}
ensures that the samples simulated using Polar codes in our algorithm
are close in distribution to those obtained
by sampling the lattice based distribution in \eqref{mc4}.

We note that the rate used in the scheme of~\cite{liu21polartit} is close to $I(\widetilde{A}\wedge \widetilde{X}\mid \widetilde{Y})$, which is shown to be close to $I(A\wedge X\mid Y)$ using similar approximations as those above. The key observation we
make is that we can decompose this rate into those corresponding to different grid points
in $\{\sigma_1^2, \dots, \sigma_r^2\}$
for $\sigma_{z}^2$,  whereby even when $\sigma_z\geq \sigma_j$ in round $j,$ the parties will agree
on the optimal auxiliary corresponding to the grid point. This auxiliary can then be subtracted from
both $X$ and $Y$, resulting in small variances for both.  Since the mean-square distance between the resulting input pairs remains the same even after this subtraction, we end-up having another instance of the same Wyner-Ziv set up. Formally, we observe the following for continuous rvs.
\begin{lemma} \label{lem1}Consider the Markov chain in \eqref{mc3}.
For every $1\leq k \leq r$ and the optimal auxiliary $A_k$ for $\sigma_k^2$ \eqref{d:aux}, the mutual information quantity  $I(A^{(k)}\wedge X|\bar{Y})$ equals to
\begin{align*}
\sum\limits_{j=1}^{k}I\left(B_{j-1}\wedge (X-A^{(j-1)}) \mid (\bar{Y}-A^{(j-1)}) \right),
\end{align*}
with $A^{(0)}=0$.
Moreover, the MMSE estimate of $X$ given $\bar{Y}$ and $A^{(k)}$ is given by
\begin{align}\label{recon1}
\hat{X}^{(k)}=A^{(k)}+\gamma\left( \bar{Y}-A^{(k)}\right),
\end{align}
where $\gamma:=\frac{\sigma_k^2\Delta(\sigma_x^2-\sigma_z^2)}{\sigma_x^2\Delta(\sigma_k^2-\sigma_z^2)+\sigma_z^2\sigma_k^2\sigma_x^2}$, under which, the distortion achieved is $\frac{\sigma_z^2\sigma_k^2\Delta}{\sigma_z^2\sigma_k^2-\Delta(\sigma_z^2-\sigma_k^2)}.$
\end{lemma}
\noindent 
{In practice, we form the estimate of $X$ by replacing $A^{(k)}$ with its decoded proxy; see Line 12, Algorithm \ref{algo2}.}
The final form of the estimate suggests that we can simply subtract $A^{(k)}$, once recovered, from $Y$ and $X$. This provides a clear justification for our
algorithm for Gaussian rvs. The main technical step is to retain these claims when we move
to discrete lattice Gaussian distribution, which we do in the manner outlined above. The following proposition characterizes the rate-loss incurred when one uses discrete lattice Gaussian  auxiliaries instead of continuous.  

\begin{proposition}[{\cite[Theorem 2]{Cong14}}] \label{flatnessprop}
Consider the Markov chains in \eqref{mc3} and \eqref{mc4} with $\Lambda=N^{-1/2}\bZ$.  Denote by $\bar{\sigma}_{k-1}$ the variance of $B_{k-1}$. Let $\widetilde{\sigma}_{\mathtt{pack}}^{(k)}:=\bar{\sigma}_{k-1}\sqrt{\frac{\mathtt{Var}(T_k+Z)}{\mathtt{Var}(Y^{(k)})}}$ and $\varepsilon_{\Lambda}(\widetilde{\sigma}_{\mathtt{pack}}^{(k)})$ be the associated flatness factor.  Then,  if $\varepsilon=\varepsilon_{\Lambda}(\widetilde{\sigma}_{\mathtt{pack}}^{(k)}) <0.5$ and $\frac{\pi\epsilon_t}{1-\epsilon_t}\leq \varepsilon,$ where 
\begin{align*}
\varepsilon_t \triangleq \begin{cases}
& \varepsilon_{\Lambda}(\bar{\sigma}_{k-1}/\sqrt{\frac{\pi}{\pi-t}}),  \, \, t\geq 1/e,\\
& (t^{-4}+1) \varepsilon_{\Lambda}(\bar{\sigma}_{k-1}/\sqrt{\frac{\pi}{\pi-t}}), \,\, 0<t< 1/e,
\end{cases}
\end{align*}
we have for all $k\in [r]$ that
\begin{align}\label{e:rate_gap1}
I(\widetilde{B}_{k-1}\wedge \widetilde{Y}^{(k)})\geq I(B_{k-1}\wedge Y^{(k)})-5\varepsilon.
\end{align}
\end{proposition}
\noindent A similar lower bound also holds in the source coding counterpart ($cf.$ \cite[Theorem 1]{liu21polartit}) for $I(\widetilde{B}_{k-1}\wedge \widetilde{X}^{(k)})$ with an equivalent variance $\widetilde{\sigma}_{\mathtt{cov}}^{(k)}=\bar{\sigma}_{k-1}\sqrt{\frac{\mathtt{var}(T_k)}{\mathtt{var}(X^{(k)})}}$ instead of $\widetilde{\sigma}_{\mathtt{pack}}^{(k)}$, i.e., 
\begin{align} \label{e:rate_gap2}
I(\widetilde{B}_{k-1}\wedge \widetilde{X}^{(k)})\geq I(B_{k-1}\wedge X^{(k)})-5\varepsilon.
\end{align}

 As evident from \eqref{e:rate_gap1} and \eqref{e:rate_gap2}, the mutual-information losses incurred due to using discrete lattice auxiliaries is atmost $5\varepsilon.$ 
As a result,  the gap to the optimal rate depends on flatness factors for the equivalent noise variances associated with $\Lambda$.  Motivated by this,  our aim is to choose a lattice $\Lambda$ such that the associated flatness factor becomes negligible.  {Also, as described above, we consider working with only $\ell=\cO(\log N)$ least significant bits.  We capture all these together formally in the following proposition.}
\begin{proposition}[{\cite[Proposition 1]{liu21polartit}}] \label{prop3}
Let $\eta=\cO(N^{-1/2})$ and $X$ be the discrete Gaussian rv over the lattice $\eta\bZ$ distributed as $\mathcal{D}_{\eta\bZ,\sigma,\bm{0}}$ {as defined in \eqref{dGauss}}. Consider an additive Gaussian noise channel having mean 0 and variance $\sigma_w^2$ with input $X$ and output $Y$.  
For $\tilde{\sigma}\triangleq\frac{\sigma\sigma_w}{\sqrt{\sigma^2+\sigma_w^2}},$ the flatness factor $\varepsilon_{\Lambda}(\tilde{\sigma})=\cO(e^{-N}).$

Moreover,  let $X_1X_2\dots$ be the binary sequence equivalent to the scaled lattice point $X/\eta$. Then, there exists an $\ell=\cO(\log N)$ such that $
\sum_{i>\ell}I(Y\wedge X_i\mid X_{1:i-1})=\cO(e^{-N})
.$
\end{proposition}
\vspace{-0.2pt}
Proposition \ref{prop3} says that
choosing the lattice to be $\eta\bZ$ ensures an exponentially small flatness factor and that, considering only the first $\ell$ least significant bits incur a capacity loss that decays exponentially in $N$.
{Recall that in order to sample from the discrete auxiliary distributions, our scheme uses Polar codes. Let  $P_{\mathbf{U}_{1:\ell}^{(k)}\mathbf{\widetilde{X}}^{(k)}\mathbf{\widetilde{Y}}^{(k)}}$ denotes the joint distribution obtained without Polar coding, i.e., not utilizing the polarization phenomenon and shared randomness at all, while simply applying randomized MAP rule ($cf.$ Line 5, Algorithm \ref{algo1}) using $\widetilde{\mathbf{X}}$ to sample 
the first $\ell=\cO(\log N)$ least significant bits for each coordinate $i\in [N]$ in all $k$ rounds.
On the other hand, let $Q_{\mathbf{U}_{1:\ell}^{(k)}\mathbf{\widetilde{X}}^{(k)}\mathbf{\widetilde{Y}}^{(k)}}$ be the joint distribution of simulated rvs obtained using the Polar coding in Algorithms \ref{algo1} and \ref{algo2}.  In the next lemma,  we show that the simulated distribution is close to the joint distribution without Polar coding in Kullback-Leibler (KL) divergence. }

\begin{lemma}\label{lem2}
For $1\leq i\leq \ell$ and $1\leq k\leq r$, the KL-divergence 
$$\mathtt{D_{KL}}(P_{\mathbf{U}_i^{(k)}|\mathbf{U}_{1:i-1}^{(k)}\mathbf{\widetilde{X}}^{(k)}}|| Q_{\mathbf{U}_i^{(k)}|\mathbf{U}_{1:i-1}^{(k)}\mathbf{\widetilde{X}}^{(k)}})=\cO(kN2^{-N^\beta}),$$
where $\beta\in (0,1/2)$ is a constant.
\end{lemma}

\noindent Using the additivity of KL-divergence and the inherent Markov chain structure
$\mathbf{U}_{i}^{(k)}\mc \mathbf{\widetilde{X}}^{(k)}\mc \mathbf{\widetilde{Y}}^{(k)}$ given $\mathbf{U}_{1:i-1}^{(k)}$, $i\in [\ell]$,  we can obtain the following corollary.
\begin{corollary}
For every $k\in [r], i\in [\ell]$, we have
\begin{align*}
\mathtt{D_{KL}}(P_{\mathbf{U}_{1:i}^{(k)}\mathbf{\widetilde{X}}^{(k)}}|| Q_{\mathbf{U}_{1:i}^{(k)}\mathbf{\widetilde{X}}^{(k)}})&=\cO(k\ell N2^{-N^\beta}), \text{and }
 \mathtt{D_{KL}}(P_{\mathbf{U}_{1:i}^{(k)}\mathbf{\widetilde{X}}^{(k)}\mathbf{\widetilde{Y}}^{(k)}}|| Q_{\mathbf{U}_{1:i}^{(k)}\mathbf{\widetilde{X}}^{(k)}\mathbf{\widetilde{Y}}^{(k)}})&=\cO(k\ell N2^{-N^\beta}),
\end{align*}
where $\beta$ is the same as in Lemma \ref{lem2}.
\label{cor2.1}
\end{corollary}

{
Next, we capture the overall performance of the proposed $r$-round WZ scheme. Recall that the unknown noise variance $\sigma_z^2\in [\sigma_0^2, \sigma_r^2],$ and the proposed scheme works only in finitely many rounds to cover this interval.  
Thus, to clearly present our ideas, we first consider the case when $\SP$ have a partial knowledge about the noise variance being one of the grid points, namely that $\sigma_z^2\in \{\sigma_1^2,\dots, \sigma_{r}^2\},$ and the set $\{\sigma_1^2,\dots, \sigma_{r}^2\}$ of possible values of the noise variance $\sigma_z^2$ is known apriori.  For this case,  the rate-distortion bound achieved by the $r$-round WZ scheme is given below.}
{\begin{theorem}\label{thm1}
Suppose that $\sigma_z=\sigma_k$ for some $k\in\{1,2,\dots, r\}.$ Then, using the scheme of Algorithm \ref{algo1} and Algorithm \ref{algo2} with the decoder for $\sigma_k,$ the reconstructed vector $\mathbf{\hat{X}}_k$ satisfies the following distortion bound
\begin{align*}
\bE\|\mathbf{X}-\mathbf{\hat{X}}^{(k)}\|^2\leq N\Delta + \cO((kN\log N)^{3/2}2^{-N^{\beta^\prime}})+\cO((kN)^2e^{-N}),
\end{align*} where $\beta^{\prime}$ is a constant in $(0,1/2)$.
Further, the total rate of communication used is roughly $\frac{1}{2}\log \frac{\sigma_{z}^2}{\Delta}+\cO(ke^{-N}).$
\end{theorem}}
We now state our result for a general case when $\sigma_z^2$ need not belong to the set $\{\sigma_1^2,\dots, \sigma_{r}^2\}.$
We remark that for this case, too, we use only finitely many rounds of communication from $\FP.$ 
{In effect, our scheme uses a $\sigma_i^2\in \{\sigma_0^2, \dots, \sigma_{r}^2\}$ that is close to $\sigma_z^2;$ the grid $\{\sigma_1^2, \dots, \sigma_{r}^2\}$ must be chosen to minimize the loss due to using $\sigma_i^2$ instead of $\sigma_z^2.$}
The following result is characterization of the universal performance of
the overall algorithm.
\begin{theorem} \label{thm2}
For every $\epsilon, \delta>0$, there exists a sufficiently large $N$ such that the scheme in Algorithm \ref{algo1}  and Algorithm \ref{algo2}
with $\ell=\mathcal{O}(\log N)$
yields an $(\epsilon, \delta)$-universal WZ code at distortion level $\Delta$ for $[\sigma_0^2, \sigma_r^2]$.
\end{theorem} Thus, the proposed scheme is our desired universal rate-optimal scheme.

\section{Proofs}
\subsection{Proof for Lemma \ref{lem1}}\label{subsecA}
Without loss of generality, let $A^{(0)}=0$ be a constant random variable. From the Markov chain in Fig. \ref{generalmc},
we have that $I(A^{(j-1)} \wedge X\mid A^{(j)},\bar{Y})=0,$ for $1\leq j\leq k-1$, which further implies $I(A^{(1)},\dots, A^{(k)}\wedge X\mid \bar{Y})=I(A^{(k)}\wedge X\mid \bar{Y})$ due the
 chain rule of mutual information. Using the chain rule again, we also have 
\begin{align*}
&I(A^{(1)},\dots, A^{(k)}\wedge X\mid \bar{Y})\\
&=\sum \limits_{j=1}^{k} I\left(A^{(j)}\wedge X \mid A^{(j-1)},\bar{Y}\right)\\
&=\sum \limits_{j=1}^{k} I\left(A^{(j)}\wedge X \mid \bar{Y}-A^{(j-1)}, A^{(j-1)}\right)\\
&=\sum \limits_{j=1}^{k} H\left(A^{(j)} \mid \bar{Y}-A^{(j-1)}, A^{(j-1)}\right) -H\left(A^{(j)} \mid \bar{Y}-A^{(j-1)}, X-A^{(j-1)}, A^{(j-1)}\right)\\
&=\sum \limits_{j=1}^{k} H\left(A^{(j)}-A^{(j-1)}\mid \bar{Y}-A^{(j-1)}\right) -H\left(A^{(j)}-A^{(j-1)}\mid \bar{Y}-A^{(j-1)}, X-A^{(j-1)}\right)\\
&=\sum \limits_{j=1}^{k}I\left((A^{(j)}-A^{(j-1)})\wedge (X-A^{(j-1)}) \mid (\bar{Y}-A^{(j-1)}) \right)\\
&=\sum\limits_{j=1}^{k}I\left(B_{j-1}\wedge (X-A^{(j-1)}) \mid (\bar{Y}-A^{(j-1)}) \right),
\end{align*}
where the second identity follows from independence of $\bar{Y}-A^{(j-1)}$ and $A^{(j-1)}$, the third identity follows from the independence of $X-A^{(j-1)}$ and $A^{(j-1)}$, and the fourth identity follows from the fact that subtracting a constant to a rv does not change its entropy, and that the difference $A^{(j)}-A^{(j-1)}$ is a Gaussian noise independent of $A^{(j-1)}.$
In addition, the MMSE estimate of $X$ given the rvs $A^{(k)}$ and $\bar{Y}$ is the conditional expectation given by
\begin{align*}
\hat{X}_k&=\bE[X\mid A^{(k)}, \bar{Y} ]\\
&=
\begin{bmatrix}
\mathtt{cov}(X,\bar{Y}) & \mathtt{cov}(X,A^{(k)})
\end{bmatrix}
\begin{bmatrix}
\mathtt{var}(\bar{Y}) &  \mathtt{cov}(\bar{Y},A^{(k)})\\
\mathtt{cov}(\bar{Y},A^{(k)}) & \mathtt{var}(\bar{Y})
\end{bmatrix}^{-1}
\begin{bmatrix}
\bar{Y}\\
A^{(k)}
\end{bmatrix}\\
&=\frac{\sigma_k^2\Delta(\sigma_x^2-\sigma_z^2)}{\sigma_x^2\Delta(\sigma_k^2-\sigma_z^2)+\sigma_z^2\sigma_k^2\sigma_x^2}\cdot \bar{Y}+\left(1-\frac{\sigma_k^2\Delta(\sigma_x^2-\sigma_z^2)}{\sigma_x^2\Delta(\sigma_k^2-\sigma_z^2)+\sigma_z^2\sigma_k^2\sigma_x^2}\right)\cdot A^{(k)}.
\end{align*}
Further, from the Markov chain in Fig. \ref{generalmc}, $\left( \bar{Y}-A^{(k)}\right)$ is independent of $A^{(k)}$ with the distribution $\mathcal{N}\left(0, \alpha_{k}\Delta_k+\frac{\sigma_x^2\sigma_{z}^2}{\sigma_x^2-\sigma_{z}^2}\right)$.
It is then easy to see that $\mathtt{var}(\hat{X}_k)=\sigma_x^2-\frac{\sigma_z^2\sigma_k^2\Delta}{\sigma_z^2\sigma_k^2-\Delta(\sigma_z^2-\sigma_k^2)}$ and that the distortion achieved $\bE[D(X,\hat{X}_k)]=\frac{\sigma_z^2\sigma_k^2\Delta}{\sigma_z^2\sigma_k^2-\Delta(\sigma_z^2-\sigma_k^2)}$.
\subsection{Proof for Lemma \ref{lem2}}\label{proof2}
First, we consider the case when the previous $k-1$ auxiliaries are recovered perfectly, by which, $\mathbf{\widetilde{X}}$ is perfectly available to be used. Using chain rule of KL-divergence, we have  
\begin{align*}
&\mathtt{D}_{\mathtt{KL}}(P_{\mathbf{U}_i^{(k)}\mid\mathbf{U}_{1:i-1}^{(k)}\mathbf{\widetilde{X}}^{(k)}}\mid\mid Q_{\mathbf{U}_i^{(k)}\mid\mathbf{U}_{1:i-1}^{(k)}\mathbf{\widetilde{X}}^{(k)}})\\
&=\sum_{j\in \FFP}\mathtt{D}_{\mathtt{KL}}(P_{\mathbf{U}_i^{(k)}(j)}\mid\mid Q_{\mathbf{U}_i^{(k)}(j)} \mid \mathbf{U}_i^{(k)}(1:j-1)\mathbf{U}_{1:i-1}^{(k)}\mathbf{\widetilde{X}}^{(k)})\\
&=\sum_{j\in \FFP}H\left( \mathbf{U}_i^{(k)}(j)\mid\mathbf{U}_i^{(k)}(1:j-1)\mathbf{U}_{1:i-1}^{(k)}\right)-H\left( \mathbf{U}_i^{(k)}(j)\mid\mathbf{U}_i^{(k)}(1:j-1)\mathbf{U}_{1:i-1}^{(k)}\mathbf{\widetilde{X}}^{(k)}\right)\\
&\leq \sum_{j\in \FFP}Z\left( \mathbf{U}_i^{(k)}(j)\mid\mathbf{U}_i^{(k)}(1:j-1)\mathbf{U}_{1:i-1}^{(k)}\right)-Z\left(\mathbf{U}_i^{(k)}(j)\mid\mathbf{U}_i^{(k)}(1:j-1)\mathbf{U}_{1:i-1}^{(k)}\mathbf{\widetilde{X}}^{(k)}\right)^2\\
&\leq 2N2^{-N^\beta},
\end{align*}
where 
the {first identity follows from the fact that the KL divergence is zero for bits indexed in information set following the randomized MAP rule in \eqref{post_prob},  the third identity follows from the $Q$ distribution induced due to decoding rule in \eqref{shared_rand}, the first inequality is due to the bounds on the conditional entropy in \eqref{BP_bounds},} and the last inequality follows from \eqref{e:bhat_cc2} in the construction of frozen set in covering code.

For the other case when we may have error in recovering previous round auxiliaries, the KL divergence $\mathtt{D_{KL}}(P_{\mathbf{U}_i^{(k)}\mid\mathbf{U}_{1:i-1}^{(k)}\mathbf{\widetilde{X}}_k}\mid\mid Q_{\mathbf{U}_i^{(k)}\mid\mathbf{U}_{1:i-1}^{(k)}\mathbf{\widetilde{X}}_k})\leq N$ {\em almost surely.} However, the expectation of decoding error probability vanishes as $O(k\cdot 2^{-N^{\beta^\prime}})$ for some $0\leq \beta^\prime\leq 0.5$.  The proof steps are similar to that of\cite[Theorem 3]{yamamoto13}) and \cite[Theorem 5]{liu21polar}. We skip the details for brevity.
Combining both these cases, the proof is completed.%
\subsection{Proof for Theorem \ref{thm1}}\label{proof3}
\paragraph{Calculation of total rate used}
The rate contribution comes from Line \ref{line9a1} of Algorithm \ref{algo1} where the difference of the frozen sets are being communicated.  Using polarization theorem of \cite{Liu19},  we have for any round $1\leq j\leq r$, 
\begin{align*}
\sum_{i=1}^\infty \frac{|\mathcal{F}_{2,i}^{(j)}|-|\mathcal{F}_{1,i}^{(j)}|}{N}&=\sum_{i=1}^\infty \frac{|\mathcal{I}_{1,i}^{(j)}|-|\mathcal{I}_{2,i}^{(j)}|}{N}\overset{N\to \infty}\longrightarrow I(\widetilde{B}_{j-1}\wedge \widetilde{X}^{(j)}\mid \widetilde{Y}^{(j)}).
\end{align*}
This implies that for an arbitrary $\varepsilon>0$,  there exists a sufficiently large  $N$ such that
\begin{align*}
\sum_{i=1}^\infty \frac{|\mathcal{F}_{2,i}^{(j)}|-|\mathcal{F}_{1,i}^{(j)}|}{N}\leq I(\widetilde{B}_{j-1}\wedge \widetilde{X}^{(j)}\mid \widetilde{Y}^{(j)})+\varepsilon.
\end{align*}
However,  in our scheme we retain only $\ell=\cO(\log N)$ levels.  Due to this, the incurred rate-loss decays exponentially as $\cO(e^{-N})$ (\textit{cf.} Proposition \ref{prop3}) and we have
\begin{align}\label{e:pol_thm}
\sum_{i=1}^\ell \frac{|\mathcal{F}_{2,i}^{(j)}|-|\mathcal{F}_{1,i}^{(j)}|}{N}\leq I(\widetilde{B}_{j-1}\wedge \widetilde{X}^{(j)}\mid \widetilde{Y}^{(j)})+\cO(e^{-N})+\varepsilon.
\end{align}
Denote by $C_{\ell,j}$ the bits communicated upto $\ell$ levels in any round $j$ is given by $C_{\ell,j}=\sum_{i=1}^\ell \frac{|\mathcal{F}_{2,i}^{(j)}|-|\mathcal{F}_{1,i}^{(j)}|}{N}$.  Further, the total rate of communication over $k$ rounds is $\sum_{j=1}^{k}C_{\ell,j}$. Using \eqref{e:pol_thm}, we bound it as 
\begin{align*}
\sum_{j=1}^{k}C_{\ell,j} 
&\leq \sum_{j=1}^{k}I(\widetilde{B}_{j-1}\wedge \widetilde{X}^{(j)}\mid \widetilde{Y}^{(j)})+\cO(ke^{-N})+k\varepsilon\\
&=\sum_{j=1}^{k}I(\widetilde{B}_{j-1}\wedge \widetilde{X}^{(j)})-I(\widetilde{B}_{j-1}\wedge \widetilde{Y}^{(j)})+\cO(ke^{-N})+k\varepsilon\\
&\leq \sum_{j=1}^{k} \left( I(B_{j-1}\wedge X^{(j)})-I(B_{j-1}\wedge Y^{(j)})\right)+\cO(ke^{-N})+\varepsilon^\prime \\
&= \sum_{j=1}^{k} I(B_{j-1}\wedge X^{(j)}\mid Y^{(j)})+\cO(ke^{-N})+\varepsilon^\prime \\
&=I(A^{(k)}\wedge X\mid \bar{Y})+\cO(ke^{-N})+\varepsilon^\prime,
\end{align*}
where the first equality is due to the underlying Markov structure, the second inequality uses  Proposition \ref{flatnessprop} and choosing $\varepsilon=\varepsilon^\prime/k$, and the last equality is due to Lemma \ref{lem1}.

\paragraph{Distortion achieved by the $r$-round WZ scheme} We calculate the distortion achieved for $\widetilde{\textbf{X}}, \widetilde{\textbf{Y}}$ under the scenarios: {\em without} and {\em with} Polar coding. 
Throughout, we denote the joint distributions induced by $P$ for the former case and by $Q$ the latter case.

\subparagraph{Distortion without Polar coding:} For this case, $\FP$ can simply apply the randomized MAP rule for every coordinate $i\in N$ using $\widetilde{\mathbf{X}}$ and sample 
the first $\ell=\cO(\log N)$ least significant bits in all rounds.
All these encoded bits at $\FP$ can then be perfectly communicated to $\SP$, which further employs an MMSE estimate to reconstruct $\widetilde{\mathbf{X}}$ within the desired distortion $\Delta$. 

Let the reconstructed source and the recovered bit vectors under the distribution $P$ be $\hat{\mathbf{X}}_{P}$ and $\{\hat{\mathbf{U}}_{i,P}^{(t)}\}_{i\geq 1, 1\leq t\leq k}$, respectively.   Denote by $\widetilde{\mathbf{A}}_{P,\infty}^{(k)}$ the auxiliary lattice point when recovered at all levels without any modulo operation, i.e., 
\[
\widetilde{\mathbf{A}}_{P,\infty}^{(k)}:=N^{-1/2}\cdot \sum_{t=1}^{k}\left((\mathbf{\hat{U}}_{1,P}^{(t)}+2\cdot \mathbf{\hat{U}}_{2,P}^{(t)}+\dots+2^{\ell-1}\cdot \mathbf{\hat{U}}_{\ell,P}^{(t)}+\dots)\mathbf{G}_N\right).
\]
However, we consider only $\ell$ levels for this case too ($cf.$ Line 9, Algorithm \ref{algo2}) to observe the corresponding auxiliary lattice point over $N^{-1/2}\bZ$ given by
\[
\widetilde{\mathbf{A}}_{P,\ell}^{(k)}:=N^{-1/2}\cdot \sum_{t=1}^{k}(\mathbf{\hat{U}}_{1,P}^{(t)}+2\cdot \mathbf{\hat{U}}_{2,P}^{(t)}+\dots+2^{\ell-1}\cdot \mathbf{\hat{U}}_{\ell,P}^{(t)})\mathbf{G}_N~\text{mod } 2^\ell\bZ.\]
 Note that the modulo operation ${~\text{ mod } 2^\ell\bZ}$ above maps inputs to the lattice points in interval $[-2^{\ell-1}, 2^{\ell-1}).$
Further, using Lemma \ref{lem1}, the final estimate used by $\SP$ is
 \[\mathbf{\hat{X}}_P=\widetilde{\mathbf{A}}_{P,\ell}^{(k)}+\gamma(\widetilde{\mathbf{Y}}-\widetilde{\mathbf{A}}_{P,\ell}^{(k)}),\]
where $\gamma$ is same as in \eqref{recon1}.
Define $\theta:=kN^{-1/2}2^{\ell-1}$ for the ease of notation. Under the joint distribution $P$, we then calculate the distortion as follows: \[\mathbb{E}_P\|\mathbf{\widetilde{X}}{-}\mathbf{\hat{X}}_P\|^2=N\mathbb{E}_P\left[(\widetilde{X}-\hat{X}_P)^2\cdot \mathbbm{1}_{\{|\widetilde{A}_{P,\infty}^{(k)}|\leq \theta\}}\right]+N\mathbb{E}_P\left[(\widetilde{X}-\hat{X}_P)^2\cdot \mathbbm{1}_{\{|\widetilde{A}_{P,\infty}^{(k)}|\geq \theta\}}\right].\]
Using the fact that $\widetilde{A}_{P,\infty}^{(k)}=\widetilde{A}_{P,\ell}^{(k)}$ whenever $|\widetilde{A}_{P,\infty}^{(k)}|\leq \theta,$ the first term in the right-hand side (RHS) simplifies to
\begin{align*}
&N\mathbb{E}_P\left[(\widetilde{X}-\hat{X}_P)^2\cdot \mathbbm{1}_{\{|\widetilde{A}_{P,\infty}^{(k)}|\leq \theta\}}\right]\\
&=N\mathbb{E}_P\left[\left(T_k-\gamma(T_k+Z')\right)^2\cdot \mathbbm{1}_{\{|\widetilde{A}_{P,\infty}^{(k)}|\leq \theta\}}\right]\\
&\leq N\frac{\sigma_z^2\sigma_k^2\Delta}{\sigma_z^2\sigma_k^2-\Delta(\sigma_z^2-\sigma_k^2)} \numberthis \label{e:exactD}\\
&=N\Delta,
\end{align*}
where the only inequality uses the fact that probability is at most 1 and the last line uses the assumption in the statement $\sigma_z=\sigma_k.$ The second term can be bounded as 
\begin{align}
&N\mathbb{E}_P\left[(\widetilde{X}-\hat{X}_P)^2\cdot \mathbbm{1}_{\{|\widetilde{A}_{P,\infty}^{(k)}|\geq\theta\}}\right]\\
&=N\mathbb{E}_P\left[(\widetilde{X}-\widetilde{A}_{P,\ell}^{(k)}-\gamma(\widetilde{Y}-\widetilde{A}_{P,\ell}^{(k)}))^2\cdot \mathbbm{1}_{\{|\widetilde{A}_{P,\infty}^{(k)}|\geq \theta\}}\right]\nonumber \\
&\leq 2N \mathbb{E}_P\left[(\widetilde{X}-\widetilde{A}_{P,\ell}^{(k)})^2\cdot \mathbbm{1}_{\{|\widetilde{A}_{P,\infty}^{(k)}|\geq \theta\}}\right]+2N\gamma^2\mathbb{E}_P\left[(\widetilde{Y}-\widetilde{A}_{P,\ell}^{(k)})^2\cdot \mathbbm{1}_{\{|\widetilde{A}_{P,\infty}^{(k)}|\geq \theta\}}\right]\label{e:p_dist},
\end{align}
where the last line uses the inequality: $(a-b)^2\leq 2(a^2+b^2).$
With regard to RHS in \eqref{e:p_dist}, the first term can be further broken as $\mathbb{E}_P\left[(\widetilde{X}-\widetilde{A}_{P,\ell}^{(k)})^2\cdot \mathbbm{1}_{\{|\widetilde{A}_{P,\infty}^{(k)}|\geq \theta\}}\right]$ can be further bounded as
\begin{align*}
&\mathbb{E}_P\left[(\widetilde{X}-\widetilde{A}_{P,\ell}^{(k)})^2\cdot \mathbbm{1}_{\{|\widetilde{A}_{P,\infty}^{(k)}|\geq \theta, |\widetilde{X}|\geq \theta\}}\right]+ \mathbb{E}_P\left[(\widetilde{X}-\widetilde{A}_{P,\ell}^{(k)})^2\cdot \mathbbm{1}_{\{|\widetilde{A}_{P,\infty}^{(k)}|\geq \theta, |\widetilde{X}|\leq \theta\}}\right]\nonumber\\
&\leq \mathbb{E}_P\left[4\widetilde{X}^2\cdot \mathbbm{1}_{\{|\widetilde{A}_{P,\infty}^{(k)}|\geq \theta, |\widetilde{X}|\geq \theta\}}\right]+\theta^2\cdot \prob(|\widetilde{A}_{P,\infty}^{(k)}|\geq \theta)\cdot \prob\left(|\widetilde{X}|\leq \theta\biggl\lvert |\widetilde{A}_{P,\infty}^{(k)}|\geq \theta\right)\nonumber\\
&\leq \sum_{|a|\geq \theta} \prob_{\widetilde{A}_{P,\infty}^{(k)}}(a) \cdot 8\int_{\theta}^\infty x^2f_{\widetilde{X}\mid \widetilde{A}_{P,\infty}^{(k)}}(x\mid a)~dx + 4\theta^2\cdot \sum_{|a|\geq \theta} \prob_{\widetilde{A}_{P,\infty}^{(k)}}(a)\cdot 1 \nonumber\\
&\leq (8c+4)\theta^2 \sum_{|a|\geq \theta} \prob_{\widetilde{A}_{P,\infty}^{(k)}}(a)\nonumber \\
&= \cO(k^2Ne^{-N^2}),\numberthis\label{e:p_dist2}
\end{align*}
where the first inequality uses the following bound
\[
(\widetilde{X}-\widetilde{A}_{P,\ell}^{(k)})^2\leq 2(|\widetilde{X}|^2+|\widetilde{A}_{P,\ell}^{(k)}|^2)\leq
\begin{cases} 
&4\widetilde{X}^2, \hbox{ for } |\widetilde{X}|\geq \theta \\
& 4\theta^2, \hbox{ for } |\widetilde{X}|\leq \theta
\end{cases}
\]
as $|\widetilde{A}_{P,\ell}^{(k)}|\leq \theta$ {\em almost surely} and 
the second inequality uses $\prob\left(|\widetilde{X}|\leq \theta \biggl\lvert |\widetilde{A}_{P,\infty}^{(k)}|\geq \theta\right)\leq 1$. and the fact that Gaussian density is an even function. 
The only integral in third line is bounded as
\begin{dmath*}
\int_{\theta}^\infty x^2f_{\widetilde{X}\mid \widetilde{A}_{P,\infty}^{(k)}}(x\mid a)~dx
=\frac{1}{\sqrt{2\pi \alpha_k\Delta_k}}\int_{\theta-a}^\infty (a^2+t^2+2at) e^{-\frac{t^2}{2\alpha_k\Delta_k}}~dt\\
\leq \left(a^2+2(2\alpha_k\Delta_k+(\theta-a)^2)+\sqrt{\frac{2}{\pi}\alpha_k\Delta_k} a \right)e^{-\frac{(\theta-a)^2}{4\alpha_k\Delta_k}}\\
\leq c\theta^2,
\end{dmath*}
where the first inequality is using the Chernoff bounds for the first component and integration by parts for the second component, and solving the integral for the third,  and the last inequality is due the fact that for $|a|\geq \theta, $ the maximum value of RHS occurs at $a=\theta$, which is a quadratic function in $\theta.$ Choosing sufficiently large values for $\ell,$ the function can be further bounded by $c\theta^2$ for some universal constant $c>0$.

The last inequality in \eqref{e:p_dist2} is due to the sub-exponential decay of discrete Gaussian distribution described below.
\begin{align*}
\sum_{|a|\geq \theta} \prob_{\widetilde{A}_{P,\infty}^{(k)}}(a)
&= \frac{2}{\sum_{\lambda \in N^{-1/2}\bZ}e^{-\frac{\lambda^2}{2\alpha_k\sigma_x^2}}} \sum_{\substack{\lambda \in N^{-1/2}\bZ, \\ \lambda\geq \theta}}e^{-\frac{\lambda^2}{2\alpha_k\sigma_x^2}} \\
&= \frac{2}{\sum_{\lambda \in N^{-1/2}\bZ}e^{-\frac{\lambda^2}{2\alpha_k\sigma_x^2}}}  \sum_{i=0}^\infty e^{-\frac{(\theta+i\cdot N^{-1/2})^2}{2\alpha_k\sigma_x^2}} \\
&\leq \frac{2}{\sum_{\lambda \in N^{-1/2}\bZ}e^{-\frac{\lambda^2}{2\alpha_k\sigma_x^2}}} \sum_{i=0}^\infty e^{-\frac{\theta^2-2\theta iN^{-1/2}}{2\alpha_k\sigma_x^2}}\\
&= \frac{2}{\sum_{\lambda \in N^{-1/2}\bZ}e^{-\frac{\lambda^2}{2\alpha_k\sigma_x^2}}}\cdot \frac{e^{-\frac{\theta^2}{2\alpha_k\sigma_x^2}}}{1-e^{-\frac{\theta N^{-1/2}}{\alpha_k\sigma_x^2}}}\\
&\leq \frac{2N^{-1/2}e^{-\frac{\theta^2}{2\alpha_k\sigma_x^2} }}{\left(1-\epsilon_{N^{-1/2}\bZ}(\sqrt{\alpha_k}\sigma_x)\right)\left(1-e^{-\frac{\theta N^{-1/2}}{\alpha_k\sigma_x^2}}\right)}\\
&=\cO(e^{-N^2}), \numberthis \label{e:DGauss_Con}
\end{align*}
where the numerator 2 in the first equality is due to the symmetry of the underlying lattice around 0,  the first inequality uses the fact that $e^{-\frac{i^2}{2\alpha_k\sigma_x^2N}}\leq 1, \forall i$,  the second inequality is due to the lower bound on normalization constant in denominator from \eqref{e:norm}, and the last line is due to $\epsilon_{N^{-1/2}\bZ}(\sqrt{\alpha_k}\sigma_x)=\cO(e^{-N})$ (see Proposition \ref{prop3}) and choosing sufficiently large $N$.

Similarly,  the second term in the RHS of \eqref{e:p_dist} can also be bounded as $\cO(k^2Ne^{-N^2}).$ Combining this bound and \eqref{e:p_dist2}, we get 
\[N\mathbb{E}_P\left[(\widetilde{X}-\hat{X}_P)^2\cdot \mathbbm{1}_{\{|\widetilde{A}_{P,\infty}^{(k)}|\geq \theta\}}\right]=\cO(k^2N^2e^{-N^2}),\]
and we have the distortion under the joint distribution $P$ obtained without Polar coding 
\begin{align}
\mathbb{E}_P\|\mathbf{\widetilde{X}}-\mathbf{\hat{X}}_P\|^2=N\Delta+\cO(k^2N^2e^{-N^2}).\label{e:dist_P}
\end{align}
However,  we must note that this case requires too many bits of communication and thus, quantization is necessary.  

\subparagraph{Distortion under Polar coding:}
Towards that, $(\FP, \SP)$ rely on the Polar coding technique, which uses shared randomness ($cf.$ Algorithm \ref{algo1} and \ref{algo2}) and exhibits much smaller communication.  
Let the reconstructed source and the recovered bit vectors under this distribution by $\hat{\mathbf{X}}_{Q}^{(k)}$ and $\{\hat{\mathbf{U}}_{i,Q}^{(t)}\}_{1\leq i\leq \ell, 1\leq t\leq k}$, respectively.   Denote by $\widetilde{\mathbf{A}}_{Q,\ell}^{(k)}$ the auxiliary lattice point over $N^{-1/2}\bZ$ when only $\ell$ levels are recovered is given by
\[
\widetilde{\mathbf{A}}_{Q,\ell}^{(k)}:=N^{-1/2}\cdot \sum_{t=1}^{k} \left((\mathbf{\hat{U}}_{1,Q}^{(t)}+2\cdot \mathbf{\hat{U}}_{2,Q}^{(t)}+\dots+2^{\ell-1}\cdot \mathbf{\hat{U}}_{\ell,Q}^{(t)})\mathbf{G}_N{}\text{ mod } 2^\ell\bZ\right).
\]
Note that the mod~${2^\ell\bZ}$ operation always maps inputs to the lattice points in interval $[-2^{\ell-1}, 2^{\ell-1}).$
Further, the final estimate used by $\SP$ is
 \[\mathbf{\hat{X}}_Q^{(k)}=\widetilde{\mathbf{A}}_{Q,\ell}^{(k)}+\gamma(\widetilde{\mathbf{Y}}-\widetilde{\mathbf{A}}_{Q,\ell}^{(k)}),\] where $\gamma$ is defined as earlier \eqref{recon1}. 
%
 Using the Minkowski's inequality\footnote{\textit{Notation alert:} In the remaining part of the proof, we use the subscripts $P$ and $Q$ to distinguish between the joint distribution under no quantization and the one simulated using Polar codes, respectively. }, we have
\begin{align}\label{e:mink1}
\sqrt{\mathbb{E}_Q\|\mathbf{\widetilde{X}}-\mathbf{\hat{X}}_Q^{(k)}\|^2}&\leq\sqrt{\mathbb{E}_P\|\mathbf{\widetilde{X}}-\mathbf{\hat{X}}_P^{(k)}\|^2} +\sqrt{\mathbb{E}_{\mu}\|\mathbf{\hat{X}}_Q^{(k)}-\mathbf{\hat{X}}_P^{(k)}\|^2},
\end{align}
for every coupling $\mu$ between $P$ and $Q$.  While we already have bound for the first term using \eqref{e:dist_P},  the second term is bounded as follows.
First note that we have
\begin{align*}
\mathbb{E}_{\mu}\|\mathbf{\hat{X}}_Q^{(k)}-\mathbf{\hat{X}}_P^{(k)}\|^2=(1-\gamma)^2\mathbb{E}_{\mu}\|\mathbf{\widetilde{A}}^{(k)}_{Q,\ell}-\mathbf{\widetilde{A}}^{(k)}_{P,\ell}\|^2. 
\end{align*}
As a result, we focus on bounding the term $\mathbb{E}_{\mu}\|\mathbf{\widetilde{A}}^{(k)}_{Q,\ell}-\mathbf{\widetilde{A}}^{(k)}_{P,\ell}\|^2$, which we do as follows. 
We have
\begin{align}
\sqrt{\mathbb{E}_{\mu}\left[\|\mathbf{\widetilde{A}}^{(k)}_{Q,\ell}-\mathbf{\widetilde{A}}^{(k)}_{P,\ell}\|^2\right]}\nonumber&\leq \sqrt{\mathbb{E}_{\mu}\left[\|\mathbf{\widetilde{A}}^{(k)}_{Q,\ell}-\mathbf{\widetilde{A}}^{(k)}_{P,\ell}\|^2\cdot \mathbbm{1}_{\{\cap_{j\in [N]}|\mathbf{\widetilde{A}}^{(k)}_{P,\infty}(j)|\leq \theta\}}\right]}\nonumber\\
&\qquad+\sqrt{\mathbb{E}_{\mu}\left[\|\mathbf{\widetilde{A}}^{(k)}_{Q,\ell}-\mathbf{\widetilde{A}}^{(k)}_{P,\ell}\|^2\cdot \mathbbm{1}_{\{\cup_{j\in [N]}|\mathbf{\widetilde{A}}^{(k)}_{P,\infty}(j)|\geq \theta\}}\right]},\label{e:q_dist1}
\end{align} 
using the Minkowski's inequality.  Further,  the first term 
in the RHS of \eqref{e:q_dist1} can be bounded as 
\begin{align*}
&\sqrt{\mathbb{E}_{\mu}\left[\left\|\mathbf{\widetilde{A}}^{(k)}_{Q,\ell}-\mathbf{\widetilde{A}}^{(k)}_{P,\ell}\right\|^2\cdot \mathbbm{1}_{\{\cap_{j\in [N]}|\mathbf{\widetilde{A}}^{(k)}_{P,\infty}(j)|\leq \theta\}}\right]}\\
&= N^{-1/2}\sqrt{\mathbb{E}_{\mu}\left[\left\|\sum_{t=1}^k\sum_{i=1}^\ell 2^{i-1}\left(\mathbf{\hat{H}}_{i,Q}^{(t)}-\mathbf{\hat{H}}_{i,P}^{(t)}\right)\right\|^2\cdot \mathbbm{1}_{\{\cap_{j\in [N]}|\mathbf{\widetilde{A}}^{(k)}_{P,\infty}(j)|\leq \theta\}}\right]}\\
&\leq N^{-1/2} 2^{\ell-1}\sum_{t=1}^k \sum_{i=1}^\ell \sqrt{\mathbb{E}_{\mu}\left[\left\|\mathbf{\hat{H}}_{i,Q}^{(t)}-\mathbf{\hat{H}}_{i,P}^{(t)}\right\|^2\cdot 1\right]}\\
&\leq N^{-1/2} 2^{\ell-1}\sum_{t=1}^k \sum_{i=1}^\ell \sqrt{\mathbb{E}_{\mu}\left[\left\|\mathbf{\hat{H}}_{i,Q}^{(t)}-\mathbf{H}_{i,Q}^{(t)}\right\|^2\right]} +\sqrt{\mathbb{E}_{\mu}\left[\left\|\mathbf{H}_{i,Q}^{(t)}-\mathbf{H}_{i,P}^{(t)}\right\|^2\right]}\\
&= N^{-1/2} 2^{\ell-1}\sum_{t=1}^k \sum_{i=1}^\ell \sqrt{\sum_{j=1}^N\mathbb{E}_{\mu}\left[\mathbbm{1}_{\{\mathbf{\hat{H}}_{i,Q}^{(t)}(j)\neq \mathbf{H}_{i,Q}^{(t)}(j)\}}\right]}+\sqrt{\sum_{j=1}^N\mathbb{E}_{\mu}\left[\mathbbm{1}_{\{\mathbf{H}_{i,Q}^{(t)}(j)\neq \mathbf{H}_{i,P}^{(t)}(j)\}}\right]}\\
&= N^{-1/2}2^{\ell-1}\sum_{t=1}^k \sum_{i=1}^\ell \sqrt{\sum_{j=1}^N\prob_{\mu} \left( \mathbf{\hat{H}}_{i,Q}^{(t)}(j)\neq \mathbf{H}_{i,Q}^{(t)}(j) \right)}+\sqrt{\sum_{j=1}^N\prob_{\mu} \left( \mathbf{H}_{i,Q}^{(t)}(j)\neq \mathbf{H}_{i,P}^{(t)}(j)\right)},\numberthis \label{e:proberr}
\end{align*}
where the first inequality uses the definition $\mathbf{\hat{H}}_i^{(t)}=\mathbf{\hat{U}}_i^{(t)}\mathbf{G}_N$ under both distributions $P$ and $Q$, and the fact that mod $2^\ell \bZ$ operation acts as an identity for the case $|\mathbf{\widetilde{A}}^{(k)}_{P,\infty}(j)|\leq \theta$, followed by the Minkowski's inequality and unity bounded indicator rvs. The second inequality uses the fact that MAP rule \eqref{post_prob2} for all the indices i.e., no Polar coding at decoder induces $P_{\mathbf{\hat{H}}_{i}^{(t)}|\mathbf{\hat{H}}_{1:i-1}^{(t)}\mathbf{\widetilde{X}}^{(t)}\mathbf{\widetilde{Y}}^{(t)}} =P_{\mathbf{H}_{i}^{(t)}|\mathbf{H}_{1:i-1}^{(t)}\mathbf{\widetilde{X}}^{(t)}\mathbf{\widetilde{Y}}^{(t)}}$ equal conditional distributions,  followed by the Minkowski's inequality.

Bounding the first term in \eqref{e:proberr} requires us to show that the decoder recovers the auxiliary with vanishing error probability.  The proof steps are similar to that of\cite[Theorem 3]{yamamoto13}) and \cite[Theorem 5]{liu21polar}. We skip the details for brevity and use the sub-exponential error bound $\cO(2^{-N^{\beta^{\prime\prime}}})$ shown for further analysis.

Towards bounding the second term in \eqref{e:proberr}, observe that \eqref{e:mink1} holds for every coupling $\mu$ and thus, we can use 
the {\em Coupling Lemma} from \cite[Lemma 3.6]{David83} to argue that there exists a coupling $\mu^\ast$ such that
\begin{align*}
&\prob_{\mu^\ast} \left( \mathbf{H}_{i,Q}^{(t)}(j)\neq \mathbf{H}_{i,P}^{(t)}(j) \right)= \mathtt{d}_{\mathtt{TV}} (P_{\mathbf{H}_i^{(t)}(j)},Q_{\mathbf{H}_i^{(t)}(j)}\mid\mathbf{H}_i^{(t)}(1:j-1)\mathbf{H}_{1:i-1}^{(t)}\mathbf{\widetilde{X}}^{(t)}\mathbf{\widetilde{Y}}^{(t)}).
\end{align*}
For such coupling, i.e., when $\mu=\mu^\ast$ in \eqref{e:proberr},  we have
 \begin{align*}
&\sqrt{\mathbb{E}_{\mu}\left[\left\|\mathbf{\widetilde{A}}^{(k)}_{Q,\ell}-\mathbf{\widetilde{A}}^{(k)}_{P,\ell}\right\|^2\cdot \mathbbm{1}_{\{\cap_{j\in [N]}|\mathbf{\widetilde{A}}^{(k)}_{P,\infty}(j)|\leq \theta\}}\right]}\\
&\leq \frac{2^{\ell-1}}{\sqrt{N}}\sum_{t=1}^k \sum_{i=1}^\ell \sqrt{\sum_{j=1}^N \mathtt{d}_{\mathtt{TV}} (P_{\mathbf{H}_i^{(t)}(j)},Q_{\mathbf{H}_i^{(t)}(j)}\mid\mathbf{H}_i^{(t)}(1:j-1)\mathbf{H}_{1:i-1}^{(t)}\mathbf{\widetilde{X}}^{(t)}\mathbf{\widetilde{Y}}^{(t)})}\\
 &\qquad \qquad\qquad \ \ \ +\cO((kN\log N)2^{-N^{\beta^{\prime\prime}}})\\
&\leq \frac{2^{\ell-1}}{\sqrt{N}}\sum_{t=1}^k \sum_{i=1}^\ell \sqrt{\sum_{j=1}^N \left(\mathtt{D}_{\mathtt{KL}} (P_{\mathbf{H}_i^{(t)}(j)},Q_{\mathbf{H}_i^{(t)}(j)}\mid\mathbf{H}_i^{(t)}(1:j-1)\mathbf{H}_{1:i-1}^{(t)}\mathbf{\widetilde{X}}^{(t)}\mathbf{\widetilde{Y}}^{(t)})\right)^{\frac{1}{2}}}\\
&\qquad \qquad \qquad \ \ \ +\cO((kN\log N)2^{-N^{\beta^{\prime\prime}}})\\
&\leq \frac{2^{\ell-1}}{\sqrt{N}}\sum_{t=1}^k \sum_{i=1}^\ell \sqrt{\sum_{j=1}^N \left(\mathtt{D_{KL}}(P_{\mathbf{H}_{1:i}^{(t)}\mathbf{\widetilde{X}}^{(t)}\mathbf{\widetilde{Y}}^{(t)}}\|Q_{\mathbf{H}_{1:i}^{(t)}\mathbf{\widetilde{X}}^{(t)}\mathbf{\widetilde{Y}}^{(t)}})\right)^{\frac{1}{2}}} +\cO((kN\log N)2^{-N^{\beta^{\prime\prime}}})\\
&= \frac{2^{\ell-1}}{\sqrt{N}}\sum_{t=1}^k \sum_{i=1}^\ell \sqrt{\sum_{j=1}^N \left(\mathtt{D_{KL}}(P_{\mathbf{U}_{1:i}^{(t)}\mathbf{\widetilde{X}}^{(t)}\mathbf{\widetilde{Y}}^{(t)}}\|Q_{\mathbf{U}_{1:i}^{(t)}\mathbf{\widetilde{X}}^{(t)}\mathbf{\widetilde{Y}}^{(t)}})\right)^{\frac{1}{2}}}+\cO((kN\log N)2^{-N^{\beta^{\prime\prime}}})\\
&= \cO((kN\log N)^{3/2}2^{-N^{\beta^\prime}}),
 \end{align*}
where 
the second inequality is the Pinsker's inequality,  the third one is using the fact that for any sequence of rvs $X_1,\dots,X_n\sim P, Y_1, \dots, Y_n\sim Q: \mathtt{D_{KL}}(P_{X_1\dots X_n}\|Q_{Y_1\dots Y_n})\geq \mathtt{D_{KL}}(P_{X_i| X_{1:i-1}}\|P_{Y_i| Y_{1:i-1}}),$ $\forall i\in [n],$
 the first identity is due to one-to-one Polar transform,  and the last line uses Lemma \ref{lem2} and $\beta^{\prime}=\beta/4$.
For the second term in \eqref{e:q_dist1}, we have 
\begin{align*}
\sqrt{\mathbb{E}_{\mu}\left[\|\mathbf{\widetilde{A}}^{(k)}_{Q,\ell}-\mathbf{\widetilde{A}}^{(k)}_{P,\ell}\|^2\cdot \mathbbm{1}_{\{\cup_{j\in [N]}|\mathbf{\widetilde{A}}^{(k)}_{P,\infty}(j)|\geq \theta\}}\right]}
&\leq \sqrt{\mathbb{E}_{\mu}\left[2(\|\mathbf{\widetilde{A}}^{(k)}_{Q,\ell}\|^2+\|\mathbf{\widetilde{A}}^{(k)}_{P,\ell}\|^2)\cdot \mathbbm{1}_{\{\cup_{j\in [N]}|\mathbf{\widetilde{A}}^{(k)}_{P,\infty}(j)|\geq \theta\}}\right]}\\
&\leq \sqrt{\mathbb{E}_{\mu}\left[4\theta^2\cdot \mathbbm{1}_{\{\cup_{j\in [N]}|\mathbf{\widetilde{A}}^{(k)}_{P,\infty}(j)|\geq \theta\}}\right]}\\
&=\cO(kNe^{-N^2}),
\end{align*}
where the first inequality uses $\|x-y\|^2\leq 2(\|x\|^2+\|y\|^2)$, the second inequality uses the fact that both $\|\mathbf{\widetilde{A}}^{(k)}_{P,\ell}\|^2$ and $\|\mathbf{\widetilde{A}}^{(k)}_{Q,\ell}\|^2$ are less than $\theta^2$ almost surely due to modulo mapping, and the last line follows by applying union bound to \eqref{e:DGauss_Con}. Plugging all the obtained upper bounds in \eqref{e:mink1}, we get 
\begin{align*}\label{e:distQ}
\mathbb{E}_Q\|\mathbf{\widetilde{X}}-\mathbf{\hat{X}}_Q^{(k)}\|^2\leq N\Delta + \cO((kN\log N)^{3/2}2^{-N^{\beta^\prime}})
+\cO((kN)^2e^{-N^2})\numberthis.
\end{align*}

Recall that all the calculations are done w.r.t input pair $(\widetilde{\mathbf{X}}, \widetilde{\mathbf{Y}})$ resulting from discrete Gaussian auxiliaries,
instead of actual input pair $(\mathbf{X}, \mathbf{Y})$. 
We now show that as the joint distributions of these pairs are close in total variational distance, the resulting distortion gap is also close.
For $\theta$ as defined earlier, we have
\begin{align*}
\mathbb{E}\|\mathbf{X}-\hat{\mathbf{X}}_Q^{(k)}\|^2
&=\iiint_{\mathbf{a},\mathbf{x},\mathbf{y}} \|\hat{\mathbf{x}}-\mathbf{x}\|^2 Q_{\mathbf{\widetilde{A}}^{(k)}_{Q,\ell}, \mathbf{X}, \mathbf{Y}}(\mathbf{a},\mathbf{x},\mathbf{y})~d\mathbf{y}d\mathbf{x}d\mathbf{a}\\
&= \iiint_{\substack{(\mathbf{a},\mathbf{x},\mathbf{y}):\forall i \text{ s.t. } \\ |\mathbf{x}(i)| \leq \theta, |\mathbf{y}(i)| \leq \theta}} \|\hat{\mathbf{x}}-\mathbf{x}\|^2 Q_{\mathbf{\widetilde{A}}^{(k)}_{Q,\ell}, \mathbf{X}, \mathbf{Y}}(\mathbf{a},\mathbf{x},\mathbf{y})~d\mathbf{y}d\mathbf{x}d\mathbf{a}\\
&\qquad +\iiint_{\substack{(\mathbf{a},\mathbf{x},\mathbf{y}):\exists i \text{ s.t. } \\ |\mathbf{x}(i)| \geq \theta \text{ or } |\mathbf{y}(i)| \geq \theta}} \|\hat{\mathbf{x}}-\mathbf{x}\|^2 Q_{\mathbf{\widetilde{A}}^{(k)}_{Q,\ell}, \mathbf{X}, \mathbf{Y}}(\mathbf{a},\mathbf{x},\mathbf{y})~d\mathbf{y}d\mathbf{x}d\mathbf{a}\\
&\leq \iiint_{\substack{(\mathbf{a},\mathbf{x},\mathbf{y}):\forall i \text{ s.t. } \\ |\mathbf{x}(i)| \leq \theta, |\mathbf{y}(i)| \leq \theta}} \|\hat{\mathbf{x}}-\mathbf{x}\|^2 (Q_{\mathbf{\widetilde{A}}^{(k)}_{Q,\ell}, \mathbf{\widetilde{X}}, \mathbf{\widetilde{Y}}}(\mathbf{a},\mathbf{x},\mathbf{y})\\
& \qquad +\left|Q_{\mathbf{\widetilde{A}}^{(k)}_{Q,\ell}, \mathbf{X}, \mathbf{Y}}(\mathbf{a},\mathbf{x},\mathbf{y})-Q_{\mathbf{\widetilde{A}}^{(k)}_{Q,\ell}, \mathbf{\widetilde{X}}, \mathbf{\widetilde{Y}}}(\mathbf{a},\mathbf{x},\mathbf{y})\right|)~d\mathbf{y}d\mathbf{x}d\mathbf{a}\\
&\qquad +\iiint_{\substack{(\mathbf{a},\mathbf{x},\mathbf{y}):\exists i \text{ s.t. }\\ |\mathbf{x}(i)| \geq \theta \text{ or } |\mathbf{y}(i)| \geq \theta}} \|\hat{\mathbf{x}}-\mathbf{x}\|^2 Q_{\mathbf{\widetilde{A}}^{(k)}_{Q,\ell}, \mathbf{X}, \mathbf{Y}}(\mathbf{a},\mathbf{x},\mathbf{y})~d\mathbf{y}d\mathbf{x}d\mathbf{a}\numberthis \label{e:integral1}
\end{align*}
 For bounding the second term in \eqref{e:integral1}, note that each $\hat{\mathbf{x}}(i)\leq \max\{\mathbf{a}(i),\mathbf{y}(i)\}\leq \theta$ a.s., which implies $\|\hat{\mathbf{x}}-\mathbf{x}\|^2\leq 4N\theta^2.$ Further, the Polar encoding operation applied to $(\mathbf{X},\mathbf{Y})$ is same as that to modified rvs $(\widetilde{\mathbf{X}},\widetilde{\mathbf{Y}})$ i.e., 
 $Q_{\mathbf{\widetilde{A}}^{(k)}_{Q,\ell}\mid \mathbf{\widetilde{X}}\mathbf{\widetilde{Y}}}(\mathbf{a}\mid \mathbf{x},\mathbf{y})=Q_{\mathbf{\widetilde{A}}^{(k)}_{Q,\ell}\mid \mathbf{X}\mathbf{Y}}(\mathbf{a}\mid \mathbf{x},\mathbf{y}).$ Thus, the first term is bounded by $8N\theta^2 \mathtt{d}_{\mathtt{TV}}(P_{\mathbf{X}\mathbf{Y}}, P_{\widetilde{\mathbf{X}}\widetilde{\mathbf{Y}}})$, which further evaluates to $\cO(N\theta^2e^{-N})$ using Lemma \ref{l:tvd} and Proposition \ref{prop3}. The first term can be bounded by $\mathbb{E}_Q\|\mathbf{\widetilde{X}}-\mathbf{\hat{X}}_Q^{(k)}\|^2$ in \eqref{e:distQ}. 
 
 Further, for the third term note that whenever there exists an $i \text{ such that } |\mathbf{y}(i)| \geq \theta, $ we have $\hat{\mathbf{x}}(i)\leq \max\{\mathbf{a}(i),\mathbf{y}(i)\}\leq |\mathbf{y}(i)|$ implies $\|\hat{\mathbf{x}}-\mathbf{x}\|^2\leq 2(\|\mathbf{y}\|^2+\|\mathbf{x}\|^2)$. As a result, we have
\begin{align*}
\iiint_{\substack{(\mathbf{a},\mathbf{x},\mathbf{y}):\exists i\\ \text{ s.t. } |\mathbf{y}(i)| \geq \theta}} \|\hat{\mathbf{x}}-\mathbf{x}\|^2 Q_{\mathbf{\widetilde{A}}^{(k)}_{Q,\ell}, \mathbf{X}, \mathbf{Y}}(\mathbf{a},\mathbf{x},\mathbf{y})~d\mathbf{y}d\mathbf{x}d\mathbf{a}&\leq \iint_{\substack{(\mathbf{x},\mathbf{y}):\exists i \text{ s.t. }\\ |\mathbf{y}(i)| \geq \theta}}2(\|y\|^2+\|\mathbf{x}\|^2) f_{\mathbf{X}\mathbf{Y}}(\mathbf{x},\mathbf{y})~d\mathbf{y}d\mathbf{x}.\numberthis\label{e:integrals}
\end{align*} 
Using Chernoff's bound and $\int_{t}^\infty y^2 dy\leq \frac{1}{\sqrt{2\pi}}te^{-\frac{t^2}{2}}{+}e^{-\frac{t^2}{2}}$, we have that
\begin{align*}
\int_{\substack{\mathbf{y}: \exists i \text{ s.t. }\\ |\mathbf{y}(i)| \geq \theta}} 2\|\mathbf{y}\|^2 f_{ \mathbf{Y}}(\mathbf{y})~d\mathbf{y}
&=\int_{\theta}^{\infty} 4y^2P_{\mathbf{Y}(i)}(y)dy+\sum_{j\neq i}\int_{y}4y^2f_{\mathbf{Y}(j)}(y)dy\cdot 2\int_{y\geq \theta} f_{\mathbf{Y}(i)}(y)dy\\
&\leq \frac{4\sigma_y^2\theta}{\sqrt{2\pi}}e^{-\frac{\theta^2}{2\sigma_y^2}}+4\sigma_y^2e^{-\frac{\theta^2}{2\sigma_y^2}}+8N\sigma_y^2e^{-\frac{\theta^2}{2\sigma_y^2}}\\
&=\cO(C_1^{-N^2}),
\end{align*}
 for a constant $C_1>1$.
 Using Chernoff's bound again, we have the second term in \eqref{e:integrals} as $\cO(e^{-\frac{\theta^2}{2\sigma_x^2}}).$
 
For the only remaining case when there exists $i$ such that $|\mathbf{x}(i)| \geq \theta \text{ and } \forall j, |\mathbf{y}(j)| \leq \theta,$ we have $\|\hat{\mathbf{x}}-\mathbf{x}\|^2\leq 4\|\mathbf{x}\|^2.$ Using similar arguments as above, we can show that for this case, too,  the integral is bounded as $\cO(C_2^{-N^2})$ for some constant $C_2>0.$
\subsection{Proof for Theorem \ref{thm2}}\label{proof4}
\noindent Recall that $\sigma_z^2$ lies in the continuum $\cI=[\sigma_0^2, \sigma_r^2]$,  but the number of rounds are finitely many.  Further, we choose the finite grid points in geometric sequence: $\sigma_i=\sigma_{i-1}2^{1/N}, 1\leq i\leq r$. This gives $k=\cO(N).$ Since,  $\SP$ knows exactly the channel variance $\sigma_z^2$,  it performs the final decoding almost surely in the round $k\in [r]$ iff $\sigma_z\in [\sigma_{k-1}, \sigma_{k})$. 
 Consider the case when $\sigma_z=\sigma_{k-1}+\varepsilon, \varepsilon>0,$ and $r$-round WZ uses $\sigma_k$ as the guess.  For this case,  the total rate of communication 
 as given by Theorem \ref{thm1} is
$\frac{1}{2}\log \frac{\sigma_k^2}{\Delta}+\cO(ke^{-N})$, which is more than the optimal rate $\frac{1}{2}\log \frac{\sigma_z^2}{\Delta}.$ Therefore, the extra rate used by the proposed universal scheme is
\begin{align*}
\sum_{j\in [k]}R_j-\frac 12 \log \frac{\sigma_{z}^2}{\Delta}&\leq \sum_{j\in [k]}R_j-\frac 12 \log \frac{\sigma_{k-1}^2}{\Delta}\\
&=\frac 12 \log \frac{\sigma_k^2}{\sigma_{k-1}^2}+ \cO(ke^{-N}) \\
&\leq \max_i \frac 12 \log \frac{\sigma_i^2}{\sigma_{i-1}^2}+\cO(ke^{-N})\\
&=\cO(1/N),
\end{align*}
where the first inequality uses $\sigma_z\geq \sigma_{k-1}$ and the last line holds for the choice of grid $\sigma_i=\sigma_{i-1}\cdot 2^{1/N}$. 

The distortion calculation goes same as in proof of Theorem \ref{thm1} earlier, except that  \eqref{e:exactD} is always bounded as 
\begin{align*}
N\mathbb{E}_P\left[(\widetilde{X}-\hat{X}_P)^2\cdot \mathbbm{1}_{\{|\widetilde{A}_{P,\infty}^{(k)}|\leq 2^{\ell-1}\}}\right]&=N\frac{\sigma_z^2\sigma_k^2\Delta}{\sigma_z^2\sigma_k^2-\Delta(\sigma_z^2-\sigma_k^2)} \\ &\leq N\Delta,
\end{align*}
because $\sigma_z< \sigma_k$, which makes the multiplicative factor less than 1. Note that the distortion bounds hold for $k=\cO(N)$ in this case.

\section*{Acknowledgement}
The author would like to thank Himanshu Tyagi for helpful discussions in formulating the problem and developing the proof ideas.  He is also grateful to Ling Liu for the discussion on Polar lattices, which helped to improve the result in Theorem \ref{thm1}.

This work is supported by Prime Minister's Research Fellowship (PMRF), Ministry of Education (MoE), India.
\bibliographystyle{IEEEtranS}
\bibliography{references}

\end{document}